\documentclass[11pt,a4paper]{article}

\usepackage{jheppub}
\usepackage[T1]{fontenc}
\usepackage{lmodern}
\usepackage{amsmath,amssymb,mathtools,mathrsfs,empheq}
\usepackage{caption}
\usepackage[dvipsnames]{xcolor}
\usepackage{stmaryrd}
\usepackage{mdframed}

\newcommand\nn{\nonumber}

\newcommand{\cN}{\mathcal{N}}
\newcommand{\bn}{\mathbf{n}}
\newcommand{\bbZ}{\mathbb{Z}}
\newcommand{\cZ}{\mathcal{Z}}
\newcommand{\cI}{\mathcal{I}}

\newcommand{\cO}{\mathcal{O}}

\newcommand{\bA}{\mathbf{A}}

\newcommand{\bF}{\mathbf{F}}
\newcommand{\tN}{\tilde{N}}
\newcommand{\tw}{\tilde{\omega}}
\newcommand{\tDelta}{\tilde{\Delta}}

\newcommand{\rH}{\rm{H}}

\newcommand{\w}{\omega}

\newcommand{\rN}{{\rm{N}}}
\newcommand{\rT}{{\rm{T}}}

\newcommand{\Q}{\mathcal{Q}}
\renewcommand{\S}{\mathcal{S}}

\newcommand{\Z}{\mathbb{Z}}

\newcommand{\I}{\mathcal{I}}

\newcommand{\N}{\mathcal{N}}

\newcommand{\para}{\paragraph{}}

\title{An M2/M5 Duality from the Giant Graviton Expansion}

\author[a]{Heng-Yu Chen,}
\author[b]{Nick Dorey,}
\author[c]{Sanefumi Moriyama,}
\author[d]{Rishi Mouland,}
\author[e]{and Canberk \c{S}anl\i}

\affiliation[a]{Department of Physics, National Taiwan University, Taipei 10617, Taiwan}
\affiliation[b]{DAMTP, Centre for Mathematical Sciences, University of Cambridge, Wilberforce Road, Cambridge CB3 0WA, U.K.}
\affiliation[c]{Department of Physics, Graduate School of Science, Osaka Metropolitan University, Sumiyoshi-ku, Osaka, Japan 558-8585}
\affiliation[d]{Abdus Salam Centre for Theoretical Physics, Imperial College London, London, SW7 2AZ, U.K. }
\affiliation[e]{CEICO, Institute of Physics of the Czech Academy of Sciences, Na Slovance 2, 182 00 Prague 8, Czech Republic}

\emailAdd{heng.yu.chen@phys.ntu.edu.tw}
\emailAdd{n.dorey@damtp.cam.ac.uk}
\emailAdd{moriyama@omu.ac.jp}
\emailAdd{r.mouland@imperial.ac.uk}
\emailAdd{sanli@fzu.cz}

\abstract{We conjecture a precise relation between the superconformal indices of two theories defined in different spacetime dimensions. The first is the three-dimensional ABJM theory describing the worldvolume of parallel M2-branes in M-theory on $\mathbb{R}^{10,1}$. The second is the $\mathcal{N}=(2,0)$ theory in six dimensions which describes the worldvolume of parallel M5-branes in the same background. As we review, the existence of such a duality is closely related to Imamura's proposal for the giant graviton expansion of the three-dimensional index. 
We check our conjecture against various results for the two indices available in the literature. Using an existing proposal of Hristov for the ABJM superconformal index, we verify our conjecture to the first three orders in an expansion around the six-dimensional Cardy limit.
} 

\begin{document}
	
\maketitle 

\section{Introduction}
\label{sec: intro}

M-theory famously has only two stable extended objects: the membrane (M2) and five-brane (M5). The membrane is electrically charged with respect to the three-form gauge potential while the five-brane couples magnetically to the same field. In fact, the two objects are electromagnetically dual and, after suitable compactification, wrapped membranes and fivebranes give rise to fundamental and solitonic strings which are related by exact dualities. The case of multiple parallel membranes or fivebranes is harder because of the new light degrees of freedom which arise when branes become coincident. In this case, each type of brane gives rise to a strongly-coupled conformal field theory (CFT) on its worldvolume. The M2 brane worldvolume is described by the ABJM $\mathcal{N}=8$ superconformal theory in three dimensions, while multiple M5 branes give rise to the $\cN=(2,0)$ theory in six dimensions. In this paper, we will investigate an interesting relation between these two CFTs which is also reminiscent of electromagnetic duality. More precisely, in the following we will conjecture such a relation for an observable which can be defined and computed exactly for both theories: the {\em superconformal index}. We will then show that our conjecture passes several tests.

\para
Each theory admits $32$ real supercharges, and hence realises a distinguished sub-sector of $\frac{1}{16}$-BPS states annihilated by a chosen Poincar\'e supercharge and its conjugate. A corresponding \textit{superconformal index} \cite{Bhattacharya:2008zy} counts these states, with an alternating sign $(-1)^F$, and a grading by the four linear combinations of bosonic symmetries commuting with the chosen supercharge. 

\para
For the M2-brane we denote this superconformal index
\begin{align}
  \I_\text{M2}(\tN;\tilde{\omega};\tilde{\Delta}_A) \ ,
\end{align}
where $\tN$ is the rank, $\tilde{\omega}$ a chemical potential for spin, and  $\tilde{\Delta}_A$, $A=1,2,3,4$ the chemical potentials for the four Cartan generators of the $SO(8)$ R-symmetry. These chemical potentials obey a linear constraint, corresponding to the fact that only 4 linear combinations of spin and R-symmetry charges commute with the chosen supercharge.

\para
Similarly, for the M5-brane, we have the superconformal index
\begin{align}
  \I_\text{M5}(N;\omega_i; \Delta_I)\ , 
\end{align}
where $N$ is the rank, $\omega_i$, $i=1,2,3$ are chemical potentials for spin, and the $\Delta_I$, $I=1,2$ are chemical potentials for the Cartan generators of the $SO(5)$ R-symmetry. Once again, these chemical potentials obey a linear constraint.

\para
The AdS/CFT correspondence is a powerful tool to study these field theoretic quantities. For instance, at leading order at large $\tN$ or $N$ and in a suitable regime of chemical potentials, $-\log \I_{\text{M2}}$ and $-\log \I_{\text{M5}}$ have been identified with the Euclidean on-shell action of known supersymmetric black hole solutions in\footnote{For further references see \cite{Cassani:2019mms} and the review \cite{Zaffaroni:2019dhb}.} AdS$_4\times S^7$ \cite{Choi:2018fdc,Cvetic:2005zi,Cassani:2019mms} and AdS$_7\times S^4$ \cite{Hosseini:2018dob,Choi:2018hmj,Nahmgoong:2019hko,Ohmori:2021dzb,Kantor:2019lfo,Bobev:2023bxl,Bobev:2025xan,Cassani:2019mms}, respectively. A sensible question is then: what more can holography teach us about these indices? A conservative approach is to consider corrections to the gravitational path integral about the black hole saddle. For instance, higher-derivative and one-loop corrections have been matched against subleading terms in the large $\tN$ expansion of $\I_\text{M2}$ \cite{Bobev:2020egg,Bobev:2021oku,Bobev:2023dwx,Bhattacharyya:2012ye,Liu:2017vbl,Liu:2017vll,Gang:2019uay,PandoZayas:2020iqr,Hristov:2021zai,David:2021eoq,Karan:2022dfy}.

\para
A rather more bold approach comes in the form of the \textit{giant graviton expansion} \cite{Gaiotto:2021xce, Imamura:2021ytr, Arai:2020uwd}. The essential idea is that $\I_{\text{M2}}$ should count $\frac{1}{16}$-BPS states of M-theory on an AdS$_4\times S^7$ background, and that these in turn are made up of BPS graviton excitations, augmented non-perturbatively by BPS excitations associated to certain brane configurations. The brane embeddings preserving the relevant supercharge are systems of M5-branes wrapping four distinguished 5-cycles $S^5\subset S^7$. 
Such configurations are generically complicated, with non-trivial overlaps between each stack of branes. A key fact we will exploit is that amongst these configurations are special classes featuring M5-branes wrapped on a single $S^5$, whose contribution to $\I_\text{M2}$ one assumes is captured by $\I_\text{M5}$. This implies a relationship between $\I_\text{M5}$ and what we will refer to as the \textit{grand canonical} superconformal
index of the M2-brane,
\begin{align}
  \Xi_\text{M2}(\mu;\tilde{\omega};\tilde{\Delta}_A):= \sum_{\tN=0}^\infty e^{\mu \tN}  \I_\text{M2}(\tN;\tilde{\omega};\tilde{\Delta}_A)\ .
\label{eq: GC SSCI}
\end{align}
We find that $\Xi_\text{M2}$ is a meromorphic function of the fugacity $\Lambda=e^{\mu}$, and that the M5-brane superconformal index can be extracted from the residue of this function at a certain pole. 
The resulting relation, which is the main claim of this paper, is
{
\setlength{\fboxsep}{6.2pt}
\begin{align}
  \boxed{
   \I_\text{M5}(N;\omega_i;\Delta_I) =-\frac{e^{\Delta_2 N}}{\I_\infty(\Delta_1;\omega_1,\omega_2,\omega_3 ,-\Delta_2)}\mathop{\text{Res}}_{\Lambda = e^{-\Delta_2 N}} \Xi_\text{M2}(\mu;\Delta_1;\omega_1,\omega_2,\omega_3 ,-\Delta_2)
  } 
    \label{eq: main claim}
\end{align}
}%
Here $\I_\infty(\tilde{\omega};\tilde{\Delta}_A)$ is an index, independent of $N$, which counts BPS graviton excitations in AdS$_4\times S^7$. It is given in (\ref{eq: I infty}). The chemical potentials appearing here obey
\begin{align}
  \Delta_1 + \Delta_2 - \omega_1 - \omega_2 - \omega_3 = 2\pi i \ .
  \label{eq: linear constraint}
\end{align}
Note that contained within the formula (\ref{eq: main claim}) is a particular identification of the chemical potentials of the M2-brane index in terms of those of the M5-brane index. We discuss this further below.

\para
It is important to note here that the proposal (\ref{eq: main claim}) relies on an analytic continuation: the idea is to compute the right-hand-side in a regime of chemical potential space $\text{Re}(\tDelta_4) = \text{Re}(-\Delta_2)>0$ in which it admits an interpretation as a convergent trace over a Hilbert space, and then analytically continue this result to a regime $\text{Re}(\Delta_2)>0$ in which the left-hand-side has such an interpretation. We will see this explicitly in the simplifying Higgs branch limit below.

\para
The purpose of this paper is to study asymptotic regimes of the parameters $\{N,\omega_i,\Delta_I\}$ in which we can verify (\ref{eq: main claim}). In all such regimes, we find an exact match between independent computations in the M2-brane and M5-brane theories. 
In addition, along the way we find that our comparatively good knowledge of $\I_\text{M2}$ will, assuming (\ref{eq: main claim}), provide new predictions for $\I_\text{M5}$ that go beyond what has been computed to date.

\para
In the remainder of this introduction, we describe the three regimes of interest: the Higgs branch limit; the leading order behaviour at large $N$; and finally, the first few orders in a six-dimensional Cardy expansion. We summarise our findings in each case, leaving details to the main text. 

\subsection{Summary of results}

Firstly, we consider (\ref{eq: main claim}) in the simplifying limit $\omega_1,\omega_2,\Delta_1\to \infty$. This coincides on the M2-brane side to the \textit{Higgs branch limit} \cite{Razamat:2014pta}, in which $\I_\text{M2}$ is given precisely by the Hilbert series of the Higgs branch. The corresponding limit of $\I_\text{M5}$ is known as the twisted limit \cite{Hayashi:2024aaf}. In this limit, we are able to exactly verify (\ref{eq: main claim}).

\para
The next most primitive check of (\ref{eq: main claim}) is to verify that the right-hand-side reproduces the expected leading large $N$ asymptotic \cite{Kantor:2019lfo,Choi:2018hmj,Hosseini:2018dob,Nahmgoong:2019hko,Bobev:2015kza,Ohmori:2021dzb}
\begin{align}\label{eq: M5 leading}
  \log \I_\text{M5}(N;\omega_i;\Delta_I) \sim -\frac{1}{24}\frac{\Delta_1^2 \Delta_2^2}{\omega_1 \omega_2 \omega_3}N^3\ .
\end{align}
To test this, we can first determine the large $\mu$ asymptotics of $\Xi_\text{M2}(\mu;\tw,\tDelta_A)$ by plugging into (\ref{eq: GC SSCI}) the leading large $\tN$ asymptotic \cite{Choi:2019zpz,Choi:2019dfu,Nian:2019pxj},
\begin{align}
  \log \I_\text{M2}(\tN;\tilde{\omega};\tilde{\Delta}_A) \sim \frac{4\sqrt{2}\,i}{3}\frac{\sqrt{\tilde{\Delta}_1\tilde{\Delta}_2\tilde{\Delta}_3\tilde{\Delta}_4}}{\tilde{\omega} }\tN^{3/2}\ ,
\end{align}
and performing a saddle-point approximation to the sum over $\tN$. From here one can extract the leading growth of the residue in (\ref{eq: main claim}), recovering precisely (\ref{eq: M5 leading}).

\para
Much less trivial is a test of (\ref{eq: main claim}) beyond this leading large $N$ behaviour. On the M2-brane side we leverage a closed form expression for $\I_\text{M2}(\tN;\tilde{\omega},\tilde{\Delta}_A)$ conjectured to be valid to all perturbative orders in $1/\tN$ \cite{Hristov:2022lcw,Hristov:2021qsw}. 
Using this proposal, we show under some mild assumptions that the right-hand-side of (\ref{eq: main claim}) admits the following asymptotic at large $N$,\footnote{Although the order $N^0$ term is unfixed, it will prove convenient to write the asymptotic in this form.}
\begin{align}
  &\log\left( -\frac{e^{\Delta_2 N}}{\I_\infty(\Delta_1;\omega_1,\omega_2,\omega_3 ,-\Delta_2)}\mathop{\text{Res}}_{\Lambda = e^{-\Delta_2 N}} \Xi_\text{M2}(\mu;\Delta_1;\omega_1,\omega_2,\omega_3 ,-\Delta_2) \right)\nn\\
  &\hspace{45mm}= \alpha(\omega_i;\Delta_I)(N^3-1) + \beta(\omega_i;\Delta_I)(N-1) + \cO\!\left(\log N\right)\ ,
  \label{eq: Hristov GC SCI}
\end{align}
where
\begin{align}
  \alpha(\omega_i;\Delta_I) 	&= -\frac{1}{24}\frac{\Delta_1^2 \Delta_2^2}{\omega_1 \omega_2 \omega_3} \ ,		\nn\\
  \beta(\omega_i;\Delta_I)	&= -\frac{1}{24\,\omega_1\omega_2\omega_3}\bigg(\, -\Delta_1\Delta_2(\Delta_1+\Delta_2 ) (\omega_1+\omega_2+\omega_3)+ \Delta_1 \Delta_2 (\omega_1+\omega_2+\omega_3)^2							\nn\\
  &\hspace{29mm}+ \left( (\Delta_1+\Delta_2)^2 -\Delta_1 \Delta_2  \right) (\omega_1\omega_2+\omega_2\omega_3+\omega_3\omega_1)							\nn\\
  &\hspace{29mm}-(\Delta_1+\Delta_2)(\omega_1+\omega_2+\omega_3)(\omega_1\omega_2+\omega_2\omega_3+\omega_3\omega_1)				\nn\\
  &\hspace{29mm}	+(\omega_1+\omega_2+\omega_3)\omega_1\omega_2\omega_3	\, \bigg)\ .
\label{eq: alpha and beta}
\end{align}
Assuming the validity of (\ref{eq: main claim}), the expression (\ref{eq: Hristov GC SCI}) provides a prediction for the large $N$ asymptotics of $\I_\text{M5}(N;\omega_i;\Delta_I)$. In particular, the leading terms can be independently identified in the 6d $(2,0)$ theory schematically as
\begin{align}
   \alpha(\omega_i;\Delta_I)(N^3-1) + \beta(\omega_i;\Delta_I)(N-1) = - \int P_8^T\ ,
\label{eq: intro TAP}
\end{align}
where the right-hand-side is (minus) the equivariant integral of the so-called `thermal anomaly polynomial'. We discuss this connection in more detail below.

\para 
To test this prediction beyond the leading $N^3$ term as above, we consider the expansion of $\I_\text{M5}(N;\omega_i;\Delta_I)$ in the Cardy regime, in which one takes $\omega_i \sim \epsilon$ small with $N$ fixed and finite. 
This expansion of 6d superconformal indices, including that of the 6d $(2,0)$ theory, was studied in \cite{Nahmgoong:2019hko,Ohmori:2021dzb}. In particular \cite{Nahmgoong:2019hko} established that in the Cardy regime one has
\begin{align}
  &\log \I_\text{M5}(N;\omega_i;\Delta_I)= \alpha(\omega_i;\Delta_I)(N^3-1) + 
  \beta(\omega_i;\Delta_I)(N-1) + \cO(\log \epsilon)\ ,
  \label{eq: Nahmgoong}
\end{align}
for the same functions $\alpha(\omega_i;\Delta_I),\beta(\omega_i;\Delta_I)$ defined in (\ref{eq: alpha and beta}). It is crucial to note here that the two terms appearing here expanded in the Cardy regime begin at order $\epsilon^{-3}$ and truncate at order $\epsilon$, and thus only the first three orders are actually fixed by \cite{Nahmgoong:2019hko}.

\para 
We see that the two results (\ref{eq: Hristov GC SCI}) and (\ref{eq: Nahmgoong}) are perfectly consistent with our key conjecture (\ref{eq: main claim}). Indeed, together they imply that we have
\begin{align}
  &\log \I_\text{M5}(N;\omega_i;\Delta_I) = \alpha(\omega_i;\Delta_I)(N^3-1) + 
  \beta(\omega_i;\Delta_I)(N-1) + \cO\left(\log \epsilon\, \log N\right)\ ,
\label{eq: combined result}
\end{align}
at large $N$ and small $\epsilon$.
In particular, we find that the fourth and fifth orders in a Cardy expansion of $\I_\text{M5}$, unfixed by \cite{Nahmgoong:2019hko}, must be precisely such that the order $N^3$ and $N$ terms match precisely the equivariant integral of the thermal anomaly polynomial as in (\ref{eq: intro TAP}).

\subsection{Outline of the paper}

In Section \ref{Sec:M2M5 relation} we demonstrate that the giant graviton expansion of $\I_\text{M2}$ due to Imamura implies our key proposal (\ref{eq: main claim}), before verifying this proposal in the Higgs branch limit. In Section \ref{Sec:3d6d SCI match} we probe (\ref{eq: main claim}) on the M2-brane side at large $\tN$, finding an exact match with the M5-brane side in the 6d Cardy limit. We provide further discussion in Section \ref{Sec:Discussion}. A handful of technical results are relegated to appendices, referred to throughout the main text.

\section{A relation between M2 and M5-brane superconformal indices}\label{Sec:M2M5 relation}

Our starting point is the superconformal index of the $U(\tN)_1\times U(\tN)_{-1}$ $\N=8$ ABJM theory \cite{Aharony:2008ug}, for which we need to set our conventions. The symmetries of this theory are described by $\frak{osp}(8|4)$. Within the conformal algebra $\frak{so}(3,2)$ we have Cartan generators: energy $\tilde{H}$ and spin $\tilde{J}$. Meanwhile we have $\frak{so}(8)$ R-symmetry Cartan generators $\tilde{Q}_A, A=1,\dots,4$. We normalise $\tilde{J},\tilde{Q}_A$ such that they take half-integer values. We finally have 32 real supercharges.

\para
We can then choose a particular Poincar\'e supercharge $\tilde{\Q}$ and its conjugate $\tilde{\S} = \tilde{\Q}^\dagger$ which impose the BPS bound
\begin{align}
  \{\tilde{\Q},\tilde{\S}\} = \tilde{H} - \tilde{J} - \frac{1}{2}(\tilde{Q}_1 + \tilde{Q}_2 + \tilde{Q}_3 + \tilde{Q}_4) \ge 0 \ .
\end{align}
The charges of $\tilde{\Q}$ under $(\tilde{H},\tilde{J},\tilde{Q}_1,\tilde{Q}_2,\tilde{Q}_3,\tilde{Q}_4)$ are $(\frac{1}{2},-\frac{1}{2},\frac{1}{2},\frac{1}{2},\frac{1}{2},\frac{1}{2})$. Let us then consider the Hilbert space trace
\begin{align}
  Z_\text{M2}(\tN;\tilde{\beta};\tilde{\omega};\tilde{\Delta}_A)=\text{Tr} \, e^{-\tilde{\beta} \{\tilde{\Q},\tilde{\S}\}}e^{-\tilde{\omega} \tilde{J} - \tilde{\Delta}_A \tilde{Q}_A}\ .
\end{align}
Generically, $Z_\text{M2}(\tN;\tilde{\beta},\tilde{\omega},\tilde{\Delta}_A)$ depends on the inverse temperature $\tilde{\beta}$ along with its five chemical potentials. However precisely when we have\footnote{More generally we can take the right-hand-side equal to $2\pi i + 4\pi i n$ for any $n\in \Z$, but such a shift can be absorbed into a $4\pi i n$ shift of any one of the chemical potentials, under which $Z_\text{M2}$ is invariant.}
\begin{align}
  \tilde{\omega} - \tilde{\Delta}_1 - \tilde{\Delta}_2 - \tilde{\Delta}_3 - \tilde{\Delta}_4 = 2\pi i\ ,
  \label{eq: linear constraint again}
\end{align}
we find the anti-commutation relation
\begin{align}
  \left\{ \tilde{\Q}, e^{-\tilde{\beta} \{\tilde{\Q},\tilde{\S}\}}e^{-\tilde{\omega} \tilde{J} - \tilde{\Delta}_A \tilde{Q}_A} \right\} = e^{-\tilde{\beta }\{\tilde{\Q},\tilde{\S}\}}\left\{ \tilde{\Q}, e^{-\tilde{\omega} \tilde{J} - \tilde{\Delta}_A \tilde{Q}_A} \right\} = 0\ ,
\end{align}
where the second equality relies on (\ref{eq: linear constraint again}). It follows that when (\ref{eq: linear constraint again}) is obeyed, $Z_\text{M2}(\tilde{\beta},\tilde{\omega},\tilde{\Delta}_A)$ becomes an index for the supercharge $\tilde{\Q}$. This means the contribution to the trace from states above the BPS bound $\{\tilde{\Q},\tilde{\S}\}\ge 0$ cancel amongst themselves, leaving only the contribution from BPS states, i.e. those with $\{\tilde{\Q},\tilde{\S}\}=0$. Accordingly $Z_{\rm{M2}}$ is now independent of $\tilde{\beta}$, and so we define superconformal index \cite{Bhattacharya:2008zy,Bhattacharya:2008bja,Kim:2009wb,Imamura:2011su,Krattenthaler:2011da,Kapustin:2011jm}
\begin{align}
  \I_\text{M2}(\tN;\tilde{\omega};\tilde{\Delta}_A) = \text{Tr} \, e^{-\tilde{\beta} \{\tilde{\Q},\tilde{\S}\}}e^{-\tilde{\omega} \tilde{J} - \tilde{\Delta}_A \tilde{Q}_A}\ ,
  \label{eq: M2 index}
\end{align}
which is a function of chemical potentials $\{\tilde{\omega},\tilde{\Delta}_A\}$ that satisfy the constraint (\ref{eq: linear constraint again}), and is independent of $\tilde{\beta}$.

\para
We then consider a proposed expansion of $\I_\text{M2}(\tN;\tilde{\omega},\tilde{\Delta}_A)$ known as the \textit{giant graviton expansion}, first put forward in this context in \cite{Arai:2020uwd}\footnote{See also \cite{Biswas:2006tj,Mandal:2006tk,Kim:2006he,Bourdier:2015wda,Arai:2019aou,Arai:2019wgv,Arai:2019xmp,Arai:2020qaj,Gaiotto:2021xce,Imamura:2021ytr,Imamura:2022aua,Murthy:2022ien} for related progress on this idea more broadly.}. The basic idea is to interpret $\I_\text{M2}$ as counting $\frac{1}{16}$-BPS excitations of M-theory on the dual geometry AdS$_4\times S^7$, where as usual the chemical potentials $\tilde{\omega}$ and $\{\tilde{\Delta}_A\}$ manifest as twists along the time direction of the $S^2\subset \text{AdS}_4$ and $S^7$, respectively. An example of such a state is  a BPS excitation of a configuration of M-branes which itself, viewed as a rigid system of defects, preserves the supercharge $\tilde{\Q}$. It is proposed then that a \textit{generic} such $\frac{1}{16}$-BPS state is obtained by dressing such a brane excitation with a gas of BPS Kaluza-Klein gravitons in AdS$_4\times S^7$. This motivates the conjectured expression 
\begin{align}
  \I_\text{M2}(\tN;\tilde{\omega}; \tilde{\Delta}_A) = \I_\infty(\tilde{\omega};\tilde{\Delta}_A)  \sum_{n_A\in \mathbb{N}_0^4} e^{-\tN \tilde{\Delta}_A n_A} \I_{(n_A)}(\tilde{\omega};\tilde{\Delta}_A)\ .
  \label{eq: GGE}
\end{align}
Let us unpack this expression. The non-negative integers $\{n_A\}$ parameterise the M-brane configurations preserving the supercharge $\Q$, which consist of four stacks of M5-branes wrapping four distinguished 5-cycles in $S^7$ (i.e. $n_1$ M5-branes wrapping the first cycle, $n_2$ wrapping the second etc.). Each 5-cycle is topologically an $S^5$. Any two 5-cycles overlap on an $S^3$, and any three overlap on an $S^1$. The prefactor $ e^{-\tN \tilde{\Delta}_An_A}$ is identified as the classical action of the $\{n_A\}$ brane configuration. Meanwhile, $\I_{(n_A)}(\tilde{\omega};\tilde{\Delta}_A)$ is interpreted as an index which counts the relevant BPS states of the worldvolume theory on the $\{n_A\}$ brane configuration. For generic $\{n_A\}$, rather little is known about $\I_{(n_A)}(\tilde{\omega};\tilde{\Delta}_A)$ beyond the large $n_A$ asymptotics suggested in \cite{Choi:2022ovw}, which were later matched against the `master volume' of certain supersymmetric geometries in AdS$_4\times S^7$ \cite{Chen:2024erz}. 
Our main interest, however will be in the case that we set three of the $n_A$ to zero, which we discuss below. Finally, the prefactor $\I_\infty(\tilde{\omega};\tilde{\Delta}_A)$ is an index counting Kaluza-Klein graviton modes in AdS$_4\times S^7$, and is given by\footnote{ 
In \cite{Arai:2020uwd}, the superconformal index $\I_\text{M2}$ is written as a manifestly meromorphic function of $\hat{q}^{1/4}$ and $\hat{u}_A^{1/2}$, where $\prod_A \hat{u}_A^{1/2} = 1$. Meanwhile in our notation, the $\I_\text{M2}$ as defined in (\ref{eq: M2 index}) is a manifestly meromorphic function of $q^{1/2}$ and $u_A^{1/2}$, which satisfy $q^{-1/2}\prod_A u_A^{1/2}=-1$ by virtue of (\ref{eq: linear constraint again}). The map between their fugacities and ours is $q^{1/2} = - \hat{q}$ and $u_A^{1/2} = \hat{q}^{1/4} \hat{u}_A^{1/2}$, where the minus sign can be traced back to their choice to separate out $(-1)^F$ explicitly, versus our choice to effectively include it in the fugacities.
 } \cite{Bhattacharya:2008zy,Arai:2020uwd}
\begin{align}
  \I_\infty(\tilde{\omega};\tilde{\Delta}_A) = \text{Pexp} \left(\frac{(1-q/u_1)(1-q/u_2)(1-q/u_3)(1-q/u_4)}{(1-u_1)(1-u_2)(1-u_3)(1-u_4)(1-q)^2}-\frac{1-q+q^2}{(1-q)^2}\right)\ ,
  \label{eq: I infty}
\end{align}
where $q=e^{-\tilde{\omega}},u_A = e^{-\tilde{\Delta}_A}$, and $\text{Pexp}$ denotes the Plethystic exponential.

\para
Let us now take (\ref{eq: GGE}) at face value, and discuss some consequences. One can reinterpret this equation in terms of the grand canonical superconformal index, defined by
\begin{align}
  \Xi_\text{M2}(\mu;\tilde{\omega};\tilde{\Delta}_A):= \sum_{\tN=0}^\infty e^{\mu \tN}  \I_\text{M2}(\tN;\tilde{\omega};\tilde{\Delta}_A)\ .
\label{eq: GC SSCI again}
\end{align}
We then have
\begin{align}
  \Xi_\text{M2}(\mu;\tilde{\omega};\tilde{\Delta}_A) &= \I_\infty(\tilde{\omega};\tilde{\Delta}_A) \sum_{\tN=0}^\infty \sum_{n_A\in \mathbb{N}_0^4} e^{\tN(\mu - \tilde{\Delta}_A n_A)} \I_{(n_A)}(\tilde{\omega}; \tilde{\Delta}_A)			\nn\\ 
  &=  \I_\infty(\tilde{\omega};\tilde{\Delta}_A) \sum_{n_A\in \mathbb{N}_0^4}\left(\sum_{\tN=0}^\infty e^{\tN(\mu - \tilde{\Delta}_A n_A)}\right)\I_{(n_A)}(\tilde{\omega}; \tilde{\Delta}_A)	\nn\\
  &= - \I_\infty (\tilde{\omega};\tilde{\Delta}_A) \sum_{n_A\in \mathbb{N}_0^4} \frac{e^{\tilde{\Delta}_A n_A } \I_{(n_A)}(\tilde{\omega};\tilde{\Delta}_A)}{\Lambda - e^{\tilde{\Delta}_A n_A}}\ ,
\label{eq: geometric sum}
\end{align}
where $\Lambda = e^{\mu}$, and we've assumed we can exchange the two sums. We see then that each giant graviton contribution $\I_{(n_A)}$ to $\I_\text{M2}$ corresponds precisely to a simple pole of $\Xi_\text{M2}$ in $\Lambda$. That is,
\begin{align}
  \mathop{\text{Res}}_{\Lambda = e^{\tilde{\Delta}_A n_A }} \Xi_\text{M2}(\mu;\tilde{\omega};\tilde{\Delta}_A) = -e^{\tilde{\Delta}_A n_A}\I_\infty (\tilde{\omega};\tilde{\Delta}_A) \I_{(n_A)} (\tilde{\omega};\tilde{\Delta}_A) \ .
\label{eq: general residue}
\end{align}
One can check this formula using the inverse formula for $\I_\text{M2}$ in terms of $\Xi_\text{M2}$, along with Cauchy's formula, as
\begin{align}
  \I_\text{M2}(\tN;\tilde{\omega};\tilde{\Delta}_A) &= \frac{1}{2\pi i} \oint_{|\Lambda|=\epsilon} \frac{d\Lambda}{\Lambda^{\tN+1}} \Xi_\text{M2}(\mu;\tilde{\omega};\tilde{\Delta}_A) 		\nn\\
  &= - \sum_{n_A \in \mathbb{N}_0^4} e^{-(\tN+1)\tilde{\Delta}_A n_A } \mathop{\text{Res}}_{\Lambda = e^{\tilde{\Delta}_A n_A }} \Xi_\text{M2}(\mu;\tilde{\omega};\tilde{\Delta}_A)	\nn\\
  &= \I_\infty(\tilde{\omega};\tilde{\Delta}_A)  \sum_{n_A\in \mathbb{N}_0^4} e^{-\tN \tilde{\Delta}_A n_A} \I_{(n_A)}(\tilde{\omega};\tilde{\Delta}_A)\ ,
\label{eq: contour}
\end{align}
recovering (\ref{eq: GGE}). Here, $\epsilon$ is chosen such that the contour contains only the pole at $\Lambda=0$, and in the second line we have evaluated this contour integral as (minus) the sum over all residues outside this contour.

\para
One should be careful however about the ranges of parameters on which these manipulations hold. The geometric sum appearing in brackets in (\ref{eq: geometric sum}) converges for all $n_A$ only if $\text{Re} [\tilde{\Delta}_A] \ge 0$ for all $I=1,\dots,4$. One also needs $|\Lambda|<|e^{\tilde{\Delta}_A  n_A}|$, with the final expression providing an analytic continuation outside this region. Meanwhile the existence of the contour $\{|\Lambda|=\epsilon\}$ in (\ref{eq: contour}) such that it only contains the pole at $\Lambda=0$ is ensured only if $\Lambda=0$ is not an accumulation point; this too requires precisely that $\text{Re}[\tilde{\Delta}_A]\ge 0$ for all $I=1,\dots,4$.

\para
We are then particularly interested in a certain special case of the general formula (\ref{eq: general residue}), in which we take $n_A=(0,0,0,N)$. The point here is that the corresponding brane configuration consists of just a single stack of $N$ M5-branes wrapping an $S^5\subset S^7$. The corresponding worldvolume theory is the 6d $U(N)$ $(2,0)$ theory, whose $\frac{1}{16}$-BPS spectrum is captured by a superconformal index $\I_\text{M5}$. Let us set our conventions to study this object. The symmetries of the theory are described by $\frak{osp}(8|4)$. Within the conformal algebra $\frak{so}(6,2)$ we have Cartan generators energy $H$ and spins $J_i$, $i=1,2,3$. Meanwhile we have $\frak{so}(5)$ Cartan generators $Q_I$, $I=1,2$. We normalise $J_i,Q_I$ such that they take half-integer values. We finally have $32$ real supercharges.

\para
We can then choose a particular Poincar\'e supercharge $\Q$ and its conjugate $\S=\Q^\dagger$ which impose the BPS bound
\begin{align}
  \{\Q,\S\} = H - J_1 - J_2 - J_3 - 2 (Q_1 +  Q_2)\ .
\end{align}
The charges of $\Q$ under $(H,J_1,J_2,J_3,Q_1,Q_2)$ are $\left(\frac{1}{2},-\frac{1}{2},-\frac{1}{2},-\frac{1}{2},\frac{1}{2},\frac{1}{2}\right)$. Then, in precisely the same way as we saw for the M2-brane theory, we can define a superconformal index \cite{Bhattacharya:2008zy,Kim:2013nva,Kim:2012qf,Kim:2012ava}
\begin{align}
  \I_\text{M5}(N;\omega_i;\Delta_I) = \text{Tr} \,e^{-\beta \{\Q,\S\}}e^{
  -\omega_1 J_1
  -\omega_2 J_2
  -\omega_3 J_3
  -\Delta_1 Q_1
  -\Delta_2 Q_2}\ ,
\end{align}
which is indeed an index for $\Q$, and is thus independent of $\beta$, provided\footnote{As we saw for the M2-brane index, more generally we can take the right-hand-side equal to $2\pi i + 4\pi i n$ for any $n\in \Z$. Once again such a shift can be absorbed into a $4\pi i n$ shift of any one of the chemical potentials, under which $\I_\text{M5}$ is invariant.}
\begin{align}
  \Delta_1 + \Delta_2 - \omega_1 - \omega_2 - \omega_3 = 2\pi i\ .
  \label{eq: M5 linear constraint}
\end{align}
We should then identify $\I_{(0,0,0,N)}(\tilde{\omega};\tilde{\Delta}_A)$ with $\I_\text{M5}(N;\omega_i;\Delta_I)$. To do so requires a map between chemical potentials. This is a purely geometric calculation, which can be approached in two (equivalent) ways. On one hand we can match charges. The M2-brane charge $\tilde{J}$ rotates the $S^2\subset \text{AdS}_4$ while the $\tilde{Q}_A$ rotate in the $S^7$. Meanwhile, the M5-brane charges $J_i$ rotate in the $S^5$ worldvolume, itself a submanifold of $S^7$, while $Q_I$ rotate in the directions transverse to the M5-branes. From here, one can read off a simple relationship between $\{\tilde{J},\tilde{Q}_A\}$ and $\{J_i, Q_I\}$. Equivalently, we can match chemical potentials directly through the geometry. Turning on $\{\tilde{\omega},\tilde{\Delta}_A\}$ corresponds to turning on off-diagonal components of the metric on AdS$_4\times S^7$ between the time direction and certain cycles in $S^2\subset \text{AdS}_4$ and $S^7$, which can then be followed to the M5-brane embedding where they correspond to twists in the metric both on and off the branes. Following either path, one arrives at \cite{Arai:2020uwd}
\begin{align}
  \I_{(0,0,0,N)}(\tilde{\omega};\tilde{\Delta}_A) = \I_\text{M5}(N;\tilde{\Delta}_1,\tilde{\Delta}_2,\tilde{\Delta}_3;\tilde{\omega},-\tilde{\Delta}_4)\ .
  \label{eq: 000N}
\end{align}
It is immediate to check that the constraint (\ref{eq: linear constraint again}) ensures (\ref{eq: M5 linear constraint}).  

\para
Finally, plugging (\ref{eq: 000N}) into (\ref{eq: general residue}) evaluated at $n_A=(0,0,0,N)$, and choosing to work in the language of M5-brane chemical potentials $\{\omega_i,\Delta_I\}$, we arrive at the key claim (\ref{eq: main claim}), which we reproduce here,
\begin{align}
\I_\text{M5}(N;\omega_i;\Delta_I) =-\frac{e^{\Delta_2 N}}{\I_\infty(\Delta_1;\omega_1,\omega_2,\omega_3 ,-\Delta_2)}\mathop{\text{Res}}_{\Lambda = e^{-\Delta_2 N}} \Xi_\text{M2}(\mu;\Delta_1;\omega_1,\omega_2,\omega_3 ,-\Delta_2)\ .
\label{eq: main claim again}
\end{align}
For the sake of complete clarity, note that the right-hand-side features the functions
$\I_\infty(\tilde{\omega};\tilde{\Delta}_A)$ and $\Xi_\text{M2}(\mu;\tilde{\omega};\tilde{\Delta}_A)$ evaluated at the chemical potential values
\begin{align}\label{eq:3d6d-ParaID-SCI}
\tilde{\omega} = \Delta_1\ ,\quad \tilde{\Delta}_1 = \omega_1\ ,\quad \tilde{\Delta}_2 = \omega_2\ ,\quad \tilde{\Delta}_3 = \omega_3\ ,\quad \tilde{\Delta}_4 = -\Delta_2\ .
\end{align}
We would now like to find independent evidence for (\ref{eq: main claim}).

\subsection{The Higgs branch limit }\label{Sec:Higgs Branch}

Let us probe the claim (\ref{eq: main claim}) in a particular simplifying limit. This is simplest to discuss by going back to the expression (\ref{eq: geometric sum}), which we can rearrange slightly into
\begin{align}
  \Xi_\text{M2}(\mu;\tilde{\omega};\tilde{\Delta}_A) &=  \I_\infty (\tilde{\omega};\tilde{\Delta}_A)\sum_{n_A\in \mathbb{N}_0^4} \frac{\I_{(n_A)}(\tilde{\omega};\tilde{\Delta}_A)}{1-\Lambda u_1^{n_1}u_2^{n_2}u_3^{n_3}u_4^{n_4} }\ ,
\label{eq: pole sum}
\end{align}
where as above we have $u_A = e^{-\tDelta_A}$ and $\Lambda=e^\mu$ .

\para
We then consider the fate of this expression if we take
\begin{align}
  \tw,\tDelta_1,\tDelta_2\to\infty\ ,
\end{align}
while holding fixed each of
\begin{align}
  \tDelta_3,\tDelta_4,(\tDelta_1+\tDelta_2-\tw)\ .
  \label{eq: fixed}
\end{align}
Taking note of the constraint (\ref{eq: linear constraint again}), we see that in this limit $\Xi_\text{M2}$ becomes a function of two independent chemical potentials $\tDelta_3,\tDelta_4$ (as well as of $\mu$). We write
\begin{align}
  \Xi_\text{M2}^\text{H}(\mu;\tDelta_3,\tDelta_4) := \lim_{\tw,\tDelta_1,\tDelta_2\to\infty} \Xi_\text{M2}(\mu;\tilde{\omega};\tilde{\Delta}_A)\ .
\end{align}
 The `H' here stands for Higgs, for this is the \textit{Higgs branch limit} \cite{Razamat:2014pta}. It is so called because in this limit $\I_\text{M2}$ coincides with the Hilbert series of the Higgs branch, a fact we will shortly utilise.
 
 \para
 Meanwhile to tackle the right-hand-side of (\ref{eq: pole sum}), let us assume that the limit
\begin{align}
  \I^\text{H}_{(n_A)}(\tDelta_3,\tDelta_4):=\lim_{\tw,\tDelta_1,\tDelta_2\to\infty}\I_{(n_A)}(\tilde{\omega};\tilde{\Delta}_A)\ ,
\end{align}
exists for all $n_A \in \mathbb{N}_0^4$, where once again we are holding fixed the quantities (\ref{eq: fixed}). Also, using (\ref{eq: I infty}) we find
\begin{align}
   \I^\text{H}_\infty (\tDelta_3,\tDelta_4) := \lim_{\tw,\tDelta_1,\tDelta_2\to\infty}\I_\infty (\tilde{\omega};\tilde{\Delta}_A) &= \text{Pexp} \left(\frac{1}{(1-u_3)(1-u_4)}-1\right)		\nn\\
   &= \prod_{\substack{n_3,n_4\ge 0 \\ (n_3,n_4)\neq (0,0)  }} \frac{1}{1-u_3^{n_3}u_4^{n_4}}\ .
\end{align}
Then in total we find that in the Higgs branch limit, (\ref{eq: pole sum}) becomes
\begin{align}
  \Xi_\text{M2}^\text{H}(\mu;\tDelta_3,\tDelta_4) =  \I^\text{H}_\infty (\tDelta_3,\tDelta_4)  \left(\sum_{n_3,n_4\ge 0} \frac{\I^\text{H}_{(0,0,n_3,n_4)}(\tDelta_3,\tDelta_4)}{1-\Lambda u_3^{n_3}u_4^{n_4} } + \sum_{\substack{n_A\in \mathbb{N}_0^4 \\ (n_1,n_2)\neq(0,0)} }\I^\text{H}_{(n_A)} (\tDelta_3,\tDelta_4)\right) \ .
  \label{eq: conjecture in Higgs limit}
\end{align}
We hence expect $\Xi_\text{M2}^\text{H}(\mu;\tDelta_3,\tDelta_4)$ to have simple poles at $\Lambda = u_3^{-n_3}u_4^{-n_4}$ for each $n_3,n_4\ge 0$, with the corresponding residues determining the (generically two-stack) giant graviton contribution $ \I^\text{H}_{(0,0,n_3,n_4)}$. Meanwhile, a constant (in $\Lambda$) remainder receives contributions from all other giant graviton configurations.

\para
The virtue of going to the Higgs branch limit is that we have more analytical control. In this limit, the M2-brane superconformal index is given by the Hilbert series that counts holomorphic functions on the Higgs branch of the ABJM theory, which at rank $\tN$ is simply the $\tN$-fold symmetric product of $\mathbb{C}^2$. This Hilbert series is weighted by fugacities $u_3,u_4$. In the grand canonical ensemble, we have simply \cite{Razamat:2014pta,Dorey:2025qht}
\begin{equation}\label{eq:Xi-M2-H}
\Xi_{\text{M2}}^{\rH}(\mu;\tDelta_{3},\tDelta_{4})= {\rm{Pexp}}\left[ \frac{\Lambda}{(1-u_3)(1-u_4)}\right] = \prod_{n_{3},n_{4}\ge 0 } \frac{1}{(1-\Lambda u_3^{n_3}u_4^{n_4})}\ .
\end{equation}
It is immediate that $\Xi_{\text{M2}}^{\rH}$ has simple poles at precisely the places predicted in (\ref{eq: conjecture in Higgs limit}). What is less obvious is that it can be arranged into the form (\ref{eq: conjecture in Higgs limit}), as generically the remainder term would be some (entire) function of $\Lambda$, while we require it to be a constant. To proceed, we utilise the Mittag-Leffler-type identity
\begin{align}
  \prod_{n,m\ge 0}\frac{1}{1-zp^n q^m} = \sum_{n,m\ge 0}\frac{1}{1-zp^n q^m}\left(\prod_{\substack{r,s \ge 0\\ (r,s)\neq (n,m)}} \frac{1}{1-p^{r-n}q^{s-m}}\right)\ ,
\label{eq: identity}
\end{align}
which we prove in Appendix \ref{app: indentity proof}. Using this, we find that $\Xi_{\text{M2}}^{\rH}(\mu;\tDelta_{3},\tDelta_{4})$ as given in (\ref{eq:Xi-M2-H}) is indeed of the form (\ref{eq: conjecture in Higgs limit}), where we identify\footnote{We define the empty product to be equal to $1$.}
\begin{align}
  \I^\text{H}_{(0,0,n_3,n_4)}(\tDelta_3,\tDelta_4) = 
  \left(\prod_{\substack{-n_3 \le r <0 \\ s\ge 0 } }  \frac{1}{1-u_3^r u_4^s}\right)
  \left(\prod_{\substack{r\ge 0\\ -n_4 \le s <0  } }  \frac{1}{1-u_3^r u_4^s}\right)
  \left(\prod_{\substack{-n_3 \le r <0\\ -n_4 \le s <0  } }  \frac{1}{1-u_3^r u_4^s}\right)\ ,
\end{align}
and\footnote{It is natural to speculate that each of the contributions $\I^\text{H}_{(n_A)} (\tDelta_3,\tDelta_4)$ with $(n_1,n_2)\neq (0,0)$ vanishes on its own, although we have no evidence for this in general.}
\begin{align}
  \sum_{\substack{n_A\in \mathbb{N}_0^4 \\ (n_1,n_2)\neq(0,0)} }\I^\text{H}_{(n_A)} (\tDelta_3,\tDelta_4) = 0\ .
\end{align}
Our central claim (\ref{eq: main claim}) then says that we should be able to identify the special case
\begin{align}
  \I^\text{H}_{(0,0,0,n_4)}(\tDelta_3,\tDelta_4) = \prod_{\substack{r\ge 0\\ -n_4 \le s <0  } }  \frac{1}{1-u_3^r u_4^s}\ ,
  \label{eq: special case}
\end{align}
with an appropriate limit of the M5-brane superconformal index.

\para
The limit in question was dubbed the \textit{twisted} limit in \cite{Hayashi:2024aaf}. Starting with the M5-brane superconformal index $\I_\text{M5}(N;\omega_i;\Delta_I)$, we take
\begin{align}
  \omega_1,\omega_2,\Delta_1\to \infty \ ,
\label{eq: twisted limit}
\end{align}
while holding fixed
\begin{align}
  \omega_3,\Delta_2,(\omega_1 + \omega_2-\Delta_1)\ .
\end{align}
The constraint (\ref{eq: M5 linear constraint}) then tells us that in taking this limit, we land on a function of two independent parameters $\omega_3,\Delta_2$ (as well as $N$), which we denote
\begin{align}
  \I_\text{M5}^\text{T}(N;\omega_3,\Delta_2) = \lim_{\omega_1,\omega_2,\Delta_1\to \infty } \I_\text{M5}(N;\omega_i;\Delta_I)\ .
\end{align}
Our central claim (\ref{eq: main claim}) then degenerates in the Higgs branch limit to the simple prediction,
\begin{align}
  \I_\text{M5}^\text{T}(N;\omega_3,\Delta_2) =  \I^\text{H}_{(0,0,0,N)}(\omega_3,-\Delta_2) =  \prod_{\substack{r\ge 0\\-N \le  s < 0  } }  \frac{1}{1-e^{-r\omega_3+s\Delta_2}}\ .
\label{eq: main caim in Higgs limit}
\end{align}
Note in particular that, as usual, we are leveraging the analytic continuation to $|u_4|>1$ provided by (\ref{eq: special case}) to write this expression.
\para 
We would finally like to verify (\ref{eq: main caim in Higgs limit}) independently on the M5-brane side. One route to do so is to utilise a formula proposed in \cite{Kim:2013nva} for $\I_\text{M5}(N;\omega_i;\Delta_I)$ determined by the localisation of an appropriate five-dimensional action on $S^1\times \mathbb{CP}^2$. This theory in turn arises from the six-dimensional $(2,0)$ theory on $S^1\times S^5$, dimensionally reduced on the Hopf fibre of $S^5$. In \cite{Dorey:2025qht}, the twisted limit (\ref{eq: twisted limit}) of this expression for $\I_\text{M5}(N;\omega_i;\Delta_I)$ was considered, and evidence was presented that it is given precisely by the formula (\ref{eq: main caim in Higgs limit}). In particular, this result was shown precisely in the `unrefined' case of $\omega_3 = \Delta_2$, while a number of checks that it continues to hold away from this special parameter subspace were performed.

\section{Large $N$ and the six-dimensional Cardy regime}\label{Sec:3d6d SCI match}

We would now like to probe (\ref{eq: main claim}) at large $N$. On the right-hand-side the behaviour at large $N$
 is encoded in the behaviour of $\Xi_\text{M2}(\mu;\tilde{\omega};\tilde{\Delta}_A)$ at large $\mu$. In detail, suppose we have an asymptotic series for the form
 \begin{align}
   \Xi_\text{M2}(\mu;\tilde{\omega};\tilde{\Delta}_A) \sim e^{G(\mu;\tilde{\omega};\tilde{\Delta}_A)} \left(1 + \dots\right)\ ,
\label{eq: asymptotic series}
\end{align}
at large $\mu$, where the growth function $G(\mu; \tilde{\omega}; \tilde{\Delta}_A)$ grows polynomially and faster than $\mu$ (it will turn out to grow like $\mu^3$). Under some mild assumption that are discussed in Appendix \ref{app: asymptotics}, it follows that as $N\to \infty$ we have
\begin{align}
  \log \left(-e^{-\tilde{\Delta}_4 N}\mathop{\text{Res}}_{\Lambda = e^{\tilde{\Delta}_4 N}} \Xi_\text{M2}(\mu;\tilde{\omega};\tilde{\Delta}_A)\right) =G(\tDelta_4 N;\tw,\tDelta) + \cO(\log N)\ .
  \label{eq: relation of asymptotics}
\end{align}
So it remains to compute the large $\mu$ behaviour of $\Xi_\text{M2}$. We can get the leading behaviour by starting with the leading large $N$ asymptotic \cite{Choi:2019zpz,Choi:2019dfu,Nian:2019pxj},
\begin{align}
  \log \I_\text{M2}(\tN;\tilde{\omega};\tilde{\Delta}_A) \sim \frac{4\sqrt{2}\,i}{3}\frac{\sqrt{\tilde{\Delta}_1\tilde{\Delta}_2\tilde{\Delta}_3\tilde{\Delta}_4}}{\tilde{\omega} }\tN^{3/2}\ .
\end{align}
From the definition (\ref{eq: GC SSCI again}) of $\Xi_\text{M2}$ we have as $\mu\to \infty$,
\begin{align}
  \Xi_\text{M2}(\mu;\tilde{\omega};\tilde{\Delta}_A)&\sim \int_0^\infty d\tN \exp\left(\mu \tN-\frac{4\sqrt{2}\,i }{3}\frac{\sqrt{\tilde{\Delta}_1\tilde{\Delta}_2\tilde{\Delta}_3\tilde{\Delta}_4}}{\tilde{\omega} }\tN^{3/2}\right) 		\nn\\
  &\sim \exp\left(-\frac{1}{24}\frac{\tilde{\omega}^2}{\tilde{\Delta}_1\tilde{\Delta}_2\tilde{\Delta}_3\tilde{\Delta}_4}\mu^3\right) \ ,	 
\end{align}
where in the first line we have employed a continuum approximation to the sum over $\tN$, which is then approximated by a saddle-point at $\tN\sim \mu^2$ which is indeed large as $\mu\to \infty$.  Then using (\ref{eq: relation of asymptotics}), we see that (\ref{eq: main claim}) implies that at leading order as $N\to \infty$,
\begin{align}
  \log \I_\text{M5}(N;\omega_i;\Delta_I) \sim -\frac{1}{24} \frac{\Delta_1^2 \Delta_2^2}{\omega_1\omega_2\omega_3} N^3\ .
\end{align}
This is precisely the expected behaviour \cite{Cassani:2019mms,Kantor:2019lfo,Choi:2018hmj,Hosseini:2018dob,Nahmgoong:2019hko,Bobev:2015kza,Ohmori:2021dzb}.

\para
We would now like to take this check of (\ref{eq: main claim}) beyond this leading large $N$ match. To do so, we first turn to the subleading in $1/\tN$ corrections to the M2-brane superconformal index.

\subsection{Perturbative in $1/\tN$ corrections to the M2-brane superconformal index}

Localisation is an invaluable tool in studying supersymetric observables in the ABJM theory. In this context, it was first used to obtain a matrix model expression for the partition function on $S^3$ \cite{Kapustin:2009kz}. The leading large $\tN$ free energy was subsequently computed to obtain the famous $\tN^{3/2}$ behaviour \cite{Drukker:2010nc}. It was soon afterwards shown that all perturbative corrections in $1/\tN$ are captured succinctly by the Airy function \cite{Fuji:2011km}.

\para 
In retrospect, the appearance of the Airy function seems very natural.
For example, the integral representation of the Airy function has a natural interpretation both in the grand canonical ensemble via the Fermi gas interpretation, and in the corresponding topological string theory \cite{Marino:2011eh}.
This interpretation goes further to all the non-perturbative corrections \cite{Hatsuda:2013oxa} which can be described in terms of the free energy of topological strings.
For these reasons, it is expected that we naturally arrive at the Airy function, even when additional chemical potentials are introduced \cite{Bobev:2025ltz}.

\para
In \cite{Hristov:2022lcw} (see also \cite{Bobev:2022jte}) interesting large $\tN$ expressions for both the squashed three-sphere partition function and the superconformal index were proposed, building on notions of gravitational blocks \cite{Hosseini:2019iad, Hristov:2021qsw} and factorisation in 4d $\N=2$ supergravity. Relevant for us is the conjecture for the large $\tN$ behaviour of the superconformal index,\footnote{The chemical potentials used in \cite{Hristov:2022lcw} are related to ours by $\tilde{\omega} = -2\pi i (\omega)_\text{theirs}$ and $ \tilde{\Delta}_A = - \pi i (\Delta_A)_\text{theirs}$. Note that \cite{Hristov:2022lcw} provides a conjectured form for the ABJM superconformal index at generic rank $k$ and with magnetic fluxes $\frak{n}$ turned on. 
We just use the special case $k=1$ with fluxes turned off here. We will comment on the extension to $k>1$ in Section \ref{Sec:Discussion}. We provide the generalisation of the manipulations performed here to generic $k$ and magnetic fluxes in Appendix \ref{App:withfluxes}. \label{Footnote:Conversion}}
\begin{align}
  \I_\text{M2}(\tN;\tilde{\omega};\tilde{\Delta}_A) \sim  e^{A(\tilde{\omega},\tilde{\Delta}_I)}\left(\text{Ai}\left[\frac{\tN-B(\tilde{\omega},\tilde{\Delta}_A)}{C(\tilde{\omega},\tilde{\Delta}_A)^{1/3}}\right]\right)^2 \left(1 + \cO\!\left(e^{-\sqrt{\tN}}\right)\right)\ ,
\label{eq: Hristov full}
\end{align}
where we define
\begin{align}\label{Def:3d-CB}
  C(\tilde{\omega},\tilde{\Delta}_I)		&= - \frac{1}{2} \frac{\tilde{\omega}^2}{\tilde{\Delta}_1\tilde{\Delta}_2\tilde{\Delta}_3\tilde{\Delta}_4}\	,\qquad B(\tilde{\omega},\tilde{\Delta}_I)		&=  \frac{\tilde{\Delta}^{(4)}-\tilde{\Delta}^{(1)}\tilde{\Delta}^{(3)}+\tilde{\omega} \tilde{\Delta}^{(1)}\tilde{\Delta}^{(2)}-\tilde{\omega}^2\tilde{\Delta}^{(2)}}{24\tilde{\Delta}_1\tilde{\Delta}_2\tilde{\Delta}_3\tilde{\Delta}_4}\ .
\end{align}
This is written in terms of the degree $n$ symmetric product
\begin{align}
  \tilde{\Delta}^{(n)} = \sum_{A_1<\dots< A_n} \tilde{\Delta}_{A_1}\dots \tilde{\Delta}_{A_n}\ .
\label{eq: symm prod}
\end{align}
In particular $B(\tw, \tDelta_I)$ is homogeneous of degree zero in chemical potentials. Meanwhile the scale factor $A(\tilde{\omega};\tilde{\Delta}_I)$ is unknown, although its behaviour in the 3d Cardy regime $\tilde{\omega}\to 0$ was constrained in \cite{Bobev:2022wem}. 

\para
To move to the grand canonical ensemble, note that we have the integral representation (abbreviating here $B=B(\tw,\tDelta_I)$ and $C=C(\tw,\tDelta_I)$)
\begin{align}
  \left(\text{Ai}\left[\frac{\tN-B}{C^{1/3}}\right]\right)^2 &= \frac{C^{1/6}}{4\pi^{3/2}i}\int_{-i\infty}^{i\infty} \frac{d\mu}{\sqrt{ \mu}} e^{-\mu \tN}\exp\left(\frac{\mu^3C}{12} +\mu B\right)		\nn\\
  &= \frac{C^{1/6}}{4\pi^{3/2}i}\int_0^{2\pi i}d\mu\, e^{-\mu \tN} \sum_{l\in \Z}\frac{1}{\sqrt{\mu - 2\pi i l}} \exp\left(\frac{(\mu-2\pi i l)^3C}{12} +(\mu-2\pi i l)B\right)	\ .
\label{eq: Airy integral}
\end{align}
Meanwhile, we have
\begin{align}
   \I_\text{M2}(\tN;\tilde{\omega};\tilde{\Delta}_A) = \frac{1}{2\pi i} \int_0^{2\pi i}d\mu\, e^{-\mu \tN} \,\Xi_\text{M2}(\mu;\tilde{\omega};\tilde{\Delta}_A)\ .
\end{align}
Using (\ref{eq: Hristov full}), we ultimately learn that (\ref{eq: Hristov full}) implies for the grand canonical superconformal index, 
\begin{align}\label{eq:logXi-M2}
  \log \Xi_\text{M2}(\mu;\tilde{\omega};\tilde{\Delta}_A) = \frac{1}{12}C(\tilde{\omega};\tilde{\Delta}_A) \mu^3 + B(\tilde{\omega},\tilde{\Delta}_A)\mu  - \frac{1}{2}\log \mu + A'(\tilde{\omega},
  \tilde{\Delta}_A) + \cO\!\left(e^{-\mu}\right) \ ,
\end{align}
where we have absorbed some constants (i.e. $\mu$-independent terms) into $A(\tilde{\omega};\tilde{\Delta}_A)$ to define $A'(\tilde{\omega};\tilde{\Delta}_A)$. This formula holds in the regime of large $\mu$ such that the $l=0$ term in (\ref{eq: Airy integral}) is dominant. Note that there are two sources of non-perturbative corrections.
One of them is the unfixed non-perturbative corrections appearing in (\ref{eq: Hristov full}), which correspond in the grand canonical ensemble to $\cO(e^{-\mu})$ corrections due to the scaling $\mu \sim N^{1/2}$ at the saddle-point.
The other is the $l\neq 0$ terms in (\ref{eq: Airy integral}) which also contribute non-perturbatively, since a decaying exponential factor appears from the expansion of cubic terms in $\mu$ \cite{Hatsuda:2012dt}.

\para 
Hence, using (\ref{eq: relation of asymptotics}), we see that
\begin{align}
  &\log\left( -\frac{e^{\Delta_2 N}}{\I_\infty(\Delta_1;\omega_1,\omega_2,\omega_3 ,-\Delta_2)}\mathop{\text{Res}}_{\Lambda = e^{-\Delta_2 N}} \Xi_\text{M2}(\mu;\Delta_1;\omega_1,\omega_2,\omega_3 ,-\Delta_2)\right)\nn\\
  &\qquad= \alpha(\omega_i;\Delta_I)(N^3-1) + \beta(\omega_i;\Delta_I)(N-1) + \cO(\log N)\ ,
\label{eq: prediction}
\end{align}
where the functions $\alpha,\beta$ are
\begin{align}
  \alpha(\omega_i;\Delta_I) 	&= \frac{1}{12}(-\Delta_2)^3 C(\Delta_1;\omega_1,\omega_2,\omega_3,-\Delta_2)		\nn\\
  \beta(\omega_i,\Delta_I)	&= (-\Delta_2) B(\Delta_1;\omega_1,\omega_2,\omega_3,-\Delta_2) \ .
  \label{eq: C and B def}
\end{align}
The explicit forms of $\alpha(\omega_i;\Delta_I),\beta(\omega_i;\Delta_I)$ are given in (\ref{eq: alpha and beta}). Note in particular that it is rather non-trivial, but nonetheless true, that both $\alpha(\omega_i;\Delta_I)$ and $\beta(\omega_i;\Delta_I)$ are indeed permutation symmetric both on the $\omega_i$ and the $\Delta_I$.
In Appendix \ref{App:superschur} we present a short discussion on the connection of $\beta(\w_i, \Delta_I)$ with supersymmetric Schur polynomials, which underlies this permutation symmetry.
\para 
We have thus established the result (\ref{eq: Hristov GC SCI}). Then, the conjectured relationship (\ref{eq: main claim}) implies a prediction for the large $N$ behaviour of $\I_\text{M5}$. Our next task is to find some supporting evidence for this prediction from the 6d $(2,0)$ theory.

\subsection{Cardy expansion of the M5-brane superconformal index } \label{subsec: Cardy}

We can now consider a particular asymptotic regime of the M5-brane superconformal index $\I_\text{M5}(N;\omega_i;\Delta_I)$ known as the \textit{Cardy-like limit}\footnote{There are in fact several Cardy-like limits of $\I_\text{M5}(N;\omega_i;\Delta_I)$. Such a limit is, from the perspective of the Euclidean path integral on $S^1\times S^5$, a sort of formal high `temperature' limit in which the index is effectively computed by a 5d EFT on $S^5$. Such a description arises precisely when $\omega_{i} \to 2\pi i m_{i}$, where the integers $m_i$ can take the inequivalent values $m_i\in \{0,1\}$ due to the $4\pi i$ periodicity of each of the $\omega_{i}$. But in fact, all states have either $J_i\in \Z$ for all $i=1,2,3$, or $J_i\in \frac{1}{2}+\Z$, by the representation theory of Spin$(6)$. It follows that there are only two inequivalent Cardy limits, labelled by $m_1+m_2+m_3$ (mod $2$). We, like \cite{Nahmgoong:2019hko,Kantor:2019lfo,Ohmori:2021dzb}, consider the case $m_1+m_2+m_3=0$ (mod $2$), which is often referred to as the Cardy limit on the `second sheet' in analogy with the analogous limit of 4d $\N=1$ indices as discussed in \cite{GonzalezLezcano:2020yeb,Cassani:2021fyv,ArabiArdehali:2021nsx}. Meanwhile, the case $m_1+m_2+m_3=1$ (mod $2$) describes the Cardy limit on the first sheet, which was studied in e.g. \cite{Chang:2017cdx,Chang:2019uag}.}. This involves taking $\omega_1,\omega_2,\omega_3\sim \epsilon$ small, while holding $(\Delta_1 - \Delta_2)$ fixed and eliminating the other combination $(\Delta_1 + \Delta_2)$ using (\ref{eq: M5 linear constraint}). In particular, each term in this expansion must be an even function of $(\Delta_1 - \Delta_2)$ by the permutation symmetries of $\I_\text{M5}$, and so for the sake of compact formulae let us define $\sigma=(\Delta_1 - \Delta_2)^2/4\pi^2$. Then, the first three orders in an expansion around small $\epsilon$, at finite $N$, were determined in \cite{Nahmgoong:2019hko}. They are given by\footnote{The chemical potentials used here are related to those of \cite{Nahmgoong:2019hko} by $(\omega_{1,2,3})_\text{there}=\omega_{1,2,3}$ and $(\Delta_L)_\text{there} = \Delta_1 - \Delta_2$,  $(\Delta_R)_\text{there} = \Delta_1 + \Delta_2$ 
\label{foot: Nahmgoong}}
\begin{align}
  &\log \I_\text{M5}(N;\omega_i;\Delta_I)  \nn\\
  &\qquad  =-\frac{\pi^2}{48\w_1 \w_2 \w_3} \Bigg(2\pi^2(\sigma+1)^2(N^3-1) 		\nn\\
  &\hspace{40mm} -4\pi i(\sigma+1)\left( (N^3 - 1)-(N-1) \right) \omega^{(1)}	\nn\\
  &\hspace{40mm} -\left((\sigma+3)(N^3-1)-4(N-1)\right)  (\omega^{(1)})^2 +	2(\sigma-3)(N-1) \omega^{(2)}	   \Bigg)	\nn\\
  &\hspace{15mm} + \cO\!\left(\log \epsilon\right)\ ,
\label{eq: M5 full Cardy}
\end{align}
where we define
\begin{align}\label{eq:omega-n}
  \omega^{(n)} = \sum_{i_1<\dots<i_n} \omega_{i_1}\dots \omega_{i_n}\ .
\end{align}
Meanwhile, we can compute the Cardy expansions of the functions appearing in the prediction (\ref{eq: prediction}) as
\begin{align}\label{eq:alpha-M5}
  \alpha(\omega_i;\Delta_I) =-\frac{\pi^2}{48\w_1 \w_2 \w_3} \Bigg(&2\pi^2(\sigma+1)^2 	-4\pi i(\sigma+1) \omega^{(1)}-(\sigma+3)  (\omega^{(1)})^2 \nn\\
  &+ \frac{i}{\pi}(\omega^{(1)})^3 + \frac{1}{8\pi^2}(\omega^{(1)})^4\Bigg)\  ,
\end{align}
and
\begin{align}\label{eq:beta-M5}
  \beta(\omega_i;\Delta_I) =-\frac{\pi^2}{48\w_1 \w_2 \w_3} \Bigg(& 4\pi i(\sigma+1) \omega^{(1)} +\left(4  (\omega^{(1)})^2 + 2(\sigma-3)\omega^{(2)}\right) \nn\\
  &+ \frac{i}{\pi}\omega^{(1)} \left(2\omega^{(2)}-(\omega^{(1)})^2 \right) + \frac{1}{2\pi^2}\omega^{(1)}\left(4\omega^{(3)}-\omega^{(1)}\omega^{(2)}\right)\Bigg)\ .
\end{align}
These are exact expressions; the Cardy expansions of $\alpha(\omega_i;\Delta_I)$ and $\beta(\omega_i;\Delta_I)$ truncate at finite order. In particular we see that $\alpha(\w_i; \Delta_I)$ begins at order $\epsilon^{-3}$ and truncates at order $\epsilon$, while $\beta(\w_i; \Delta_I)$ begins at order $\epsilon^{-2}$ and truncates at order $\epsilon$. In each expression, we have written the terms up to and including those of order $\epsilon^{-1}$ on the first line, and then subleading terms on the second line.
\para
It is straightforward then to see that (\ref{eq: M5 full Cardy}) can be re-expressed in the form
\begin{align}
  \log \I_\text{M5}(N;\omega_i;\Delta_I) = \alpha(\omega_i;\Delta_I)(N^3-1) + \beta(\omega_i;\Delta_I)(N-1) + \cO(\log \epsilon)\ .
\end{align}
Thus, we establish the expression (\ref{eq: Nahmgoong}).

\para
We learn that in a combined large $N$ and Cardy expansion, our two results (\ref{eq: Hristov GC SCI}) and (\ref{eq: Nahmgoong}) match to the first three orders in a Cardy expansion. But indeed we get more than this: the M2-brane result (\ref{eq: Hristov GC SCI}) makes a prediction for the coefficients of the $N^3$ and $N$ terms to \textit{all orders} in the Cardy expansion. In particuar, the fourth and fifth orders are non-zero, since the expansions of $\alpha(\omega_i;\Delta_I),\beta(\omega_i;\Delta_I)$ do not truncate until order $\epsilon$. A natural question, then, is: is there an interpretation of these next two orders from the perspective of the M5-brane theory? As we will see next, the answer is \textit{yes}.

\subsection{Relation to the thermal anomaly polynomial}\label{subsec: anomaly}

So, we are looking for an M5-brane interpretation of our functions $\alpha(\omega_i;\Delta_I),\beta(\omega_i;\Delta_I)$. Our starting point must therefore be some object, computable purely in the 6d $U(N)$ $(2,0)$ theory, which agrees to the first three orders in the Cardy expansion with $\log \I_\text{M5}$. Precisely such an object was already identified in \cite{Nahmgoong:2019hko}.

\para
Recall that \cite{Nahmgoong:2019hko} determined only the first three orders of $\log \I_\text{M5}$ in the Cardy expansion. However, it was noted there that these first three orders matched with the first three orders in a Cardy expansion of another object\footnote{A related but distinct object is the supersymmetric Casimir energy, which governs the low temperature behaviour of the supersymmetric partition function, and which was conjectured in \cite{Bobev:2015kza} to be given by (minus) the equivariant integral of the anomaly polynomial $P_8$. The relation between these two anomaly polynomials $P_8^T$ and $P_8$, and thus the contribution that $-\int P_8^T$ receives from the supersymmetric Casimir energy, was discussed in \cite{Ohmori:2021dzb}.}: the equivariant integral of the `thermal anomaly polynomial' $P_8^T$, which is itself derived from the usual anomaly polynomial $P_8$ following \cite{Jensen:2013rga,Ohmori:2021dzb}. Schematically,
\begin{align}
   \log \I_\text{M5}(N;\omega_i;\Delta_I) = - \int_{(\omega_i;\Delta_I)} P_8^T + \cO(\log \epsilon)\ ,
\label{eq: Nahmgoong anomaly}
\end{align}
in the Cardy limit $\epsilon\to 0$. The subscript on the integral reminds us to fix the equivariant parameters appearing in \cite{Nahmgoong:2019hko} appropriately. The value for this equivariant integral is given in equation (4.6) of \cite{Nahmgoong:2019hko}, the derivation of which is briefly reviewed in Appendix \ref{App:superschur}. Using this expression we find precisely
\begin{align}\label{eq:TSCE-P8}
  - \int_{(\omega_i;\Delta_I)} P_8^T &= \alpha(\omega_i,\Delta_I)(N^3-1) + \beta(\omega_i,\Delta_I)(N-1) 		\ .
\end{align}
In particular, the Cardy expansion of the equivariant integral truncates at order $\epsilon$. We see therefore that the order $\epsilon^0$ and $\epsilon^1$ corrections to the Cardy formula of \cite{Nahmgoong:2019hko} as implied by the M2-brane result (\ref{eq: Hristov GC SCI}) match precisely with equivariant integral of the thermal anomaly polynomial.
\para 
We learn that the validity of the central claim (\ref{eq: main claim}) implies a strengthened version of (\ref{eq: Nahmgoong anomaly}), which is that in the Cardy limit $\epsilon\to 0$ and at large $N$, we have
\begin{align}
   \log \I_\text{M5}(N;\omega_i;\Delta_I)= - \int_{(\omega_i;\Delta_I)} P_8^T + \cO\left(\log \epsilon\, \log N\right)\ .
\label{eq: strengthened Cardy}
\end{align}
Let us finally note that this is not the first time that a strengthened form of the result (\ref{eq: Nahmgoong anomaly}) has been proposed. The paper \cite{Ohmori:2021dzb} studied in detail the Cardy limit on the second sheet of 4d $\N=1$ indices, and in doing so arrived at a Cardy formula in terms of an analogous thermal anomaly polynomial. It is argued there that this formula governs the Cardy behaviour of the index up to corrections that are \textit{exponentially suppressed} in this regime. Based on these results, it was proposed that a similar formula might hold in six dimensions. In our language, for the 6d $U(N)$ $(2,0)$ theory, this would mean
\begin{align}
   \log \I_\text{M5}(N;\omega_i;\Delta_I) \stackrel{?}{=} - \int_{(\omega_i;\Delta_I)} P_8^T + \cO\!\left( e^{-1/\epsilon}\right)  \ .
\label{eq: Ohmori}
\end{align}
We see that this is perfectly consistent with our M2-brane result (\ref{eq: prediction}) and the key claim (\ref{eq: main claim}), and indeed implies that the corrections appearing in (\ref{eq: strengthened Cardy}) should be of order $e^{-1/\epsilon} \log N$.

\section{Discussion}\label{Sec:Discussion}

In this paper, we have leveraged the giant graviton expansion---a tool born of AdS/CFT---to arrive at a reasonably simple conjecture (\ref{eq: main claim}) relating purely field-theoretic quantities. This conjecture relates the $\frac{1}{16}$-BPS spectra of a three-dimensional $\N=8$ SCFT with that of a six-dimensional $\N=(2,0)$ SCFT, in a quantitative way.

\para
We have then performed a series of checks of this relationship. We first verified it explicitly in the simplifying Higgs branch limit of the three-dimensional theory, corresponding to the twisted limit of the six-dimensional theory. Next, we straightforwardly checked that the leading large $\tN$ behaviour of the M2-brane superconformal index $\I_\text{M2}(\tN;\tw,\tDelta_A)$ reproduces the expected leading large $N$ behaviour of the M5-brane superconformal index $\I_\text{M5}(N;\omega_i;\Delta_I)$. Finally, we explored subleading in $1/\tN$ corrections to $\I_\text{M2}(\tN;\tw,\tDelta_A)$, extracting from them a prediction (\ref{eq: Hristov GC SCI}) for $\log \I_\text{M5}(N;\omega_i;\Delta_I)$ up to order $N^1$ and at generic chemical potentials. This prediction was shown to match precisely with an independent computation (\ref{eq: Nahmgoong}) valid to the first three orders in the six-dimensional Cardy expansion.

\para
Let us conclude with some comments and future directions.

\begin{itemize}
  \item The conjecture (\ref{eq: Hristov full}) for the large $\tN$ behaviour of $\I_\text{M2}$ is really rather strong, for it predicts (with the exception of the constant term) all perturbative contributions to $\log \I_\text{M2}$. In this paper we have barely scratched the surface of what this proposal implies for the large $N$ expansion of $\I_\text{M5}$ if we are to trust (\ref{eq: main claim}). To be precise, all that we have used is the first two non-trivial orders in an expansion of (\ref{eq: Hristov full}) at large $\tN$, i.e.\footnote{Note that the coefficient of the order $\tN^{1/2}$ term here was independently verified in \cite{Bobev:2022wem} at leading order in the three-dimensional Cardy limit $\tw\to 0$.}
  \begin{align}
  \log \I_\text{M2}(\tN;\tw;\tDelta_A) = \left(- \frac{4}{3C(\tw;\tDelta_A)^{1/2}} \right) \tN^{3/2} + \left(\frac{2B(\tw;\tDelta_A)}{C(\tw;\tDelta_A)^{1/2}} \right) \tN^{1/2} + \cO(\log \tN) \ .
\end{align}
Said another way, the order $N^3$ and $N$ terms of $\log \I_\text{M5}(N;\omega_i;\Delta_I)$ are sensitive only to the order $\tN^{3/2}$ and $\tN^{1/2}$ terms of $\log \I_\text{M2}(\tN;\tw;\tDelta_A)$.

It is obviously then desirable to use the subleading data contained in (\ref{eq: Hristov GC SCI}) at orders beyond $\tN^{1/2}$ to provide further predictions for $\log \I_\text{M5}(N;\omega_i;\Delta_I)$ at orders beyond order $N$. The first obvious target is the coefficient of the $\log N$ term\footnote{This coefficient was conjectured to be zero up to corrections that are non-perturbatively small in the Cardy limit in \cite{Ohmori:2021dzb}. However it appears feasible that the analysis there may need to be corrected by a term $\log |G| = \log |\Z_N| = \log N$ corresponding to the spontaneously broken $\Z_N$ 2-form symmetry of the 6d $U(N)$ $(2,0)$ theory, in analogy with the $\log |G|$ term appearing in 4d $\N=1$ Cardy formulae arising from a spontaneously broken 1-form symmetry $G$ \cite{Cassani:2021fyv}. We thank Luigi Tizzano for pointing this out to us.}, which one expects to match holographically against a counting of zero modes in the relevant back hole background \cite{Bobev:2023bxl}, as has been studied in much detail in AdS$_4$ \cite{Bhattacharyya:2012ye,Liu:2017vbl,Liu:2017vll,Gang:2019uay,PandoZayas:2020iqr,Hristov:2021zai,David:2021eoq,Karan:2022dfy,Bobev:2023dwx}.

This paper provides a roadmap for extracting such higher order corrections starting with the conjecture (\ref{eq: Hristov full}), in the form of corrections to the saddle-point computation discussed in Appendix \ref{app: asymptotics}. 

\item Our result (\ref{eq: combined result}) implies that the coefficient of $N$ in the large $N$ expansion of $\log \I_\text{M5}(N;\omega_i;\Delta_I)$ is precisely $\beta(\omega_i;\Delta_I)$. Previous results, consistent with ours, only predicted this coefficient to the first three orders in the six-dimensional Cardy expansion \cite{Nahmgoong:2019hko}, which was later strengthened to a conjecture of its form to all perturbative orders in the Cardy expansion in \cite{Ohmori:2021dzb}.

Now with evidence in hand that the coefficient is precisely $\beta(\omega_i;\Delta_I)$, the time is ripe for this coefficient to be matched holographically. In analogy with what has been achieved in AdS$_4$ \cite{Bobev:2021oku,Bobev:2020egg} and AdS$_5$ \cite{Bobev:2022bjm}, this coefficient should correspond to the correction to the Euclidean on-shell action of the relevant AdS$_7$ BPS black hole \cite{Bobev:2023bxl,Bobev:2025xan} when one includes higher derivative corrections to 7d gauged supergravity.

\item Our focus in this paper was the $U(\tN)_k \times U(\tN)_{-k}$ ABJM theory at level $k=1$, and the giant graviton expansion (\ref{eq: GGE}) of its superconformal index. This is straightforwardly generalised to $k>1$ \cite{Arai:2020uwd}. The dual geometry is AdS$_4\times S^7/\mathbb{Z}_k$, with giant gravitons now wrapping a smooth orbifold $S^5/\Z_k \subset S^7 /\Z_k$.

The conjecture (\ref{eq: Hristov full}) generalises simply as \cite{Hristov:2022lcw}
\begin{equation}\label{eq:3d-BC-k}
C(\tw, \tDelta_A) \to \frac{1}{k}C(\tw, \tDelta_A)\ , \quad B(\tw, \tDelta_A) \to \frac{1}{k} B(\tw, \tDelta_A)+\frac{1}{24}\left(k-\frac{1}{k}\right)\ .
\end{equation}
One can understand the rescaling by $1/k$ as arising from the relative volume of $S^7/\Z_k$ versus $S^7$. Meanwhile, the shift of $\frac{1}{24}(k - \frac{1}{k})$ can be attributed to an anomalous M2-brane charge arising from the fixed point of $\mathbb{C}^4/\Z_k$ probed by the $\tN$ M2-branes, via a gravitational anomaly term \cite{Bergman:2009zh}.

By extracting a suitable residue, we can obtain a prediction for the superconformal index of the 6d $U(N)$ $(2,0)$ theory on\footnote{One has to be precise about how this orbifold is defined. In particular since the $\Z_k$ orbifold acts simultaneously in the four planes of the $\mathbb{C}^4$ in which $S^7$ is embedded, it follows that the $\Z_k$ orbifold of the M5-brane theory acts simultaneously in the three planes of $\mathbb{C}^3$ in which the spatial $S^5$ is embedded as well as along the R-symmetry direction generated by $-Q_2$. This explains the lack of symmetry of the (\ref{eq: S5/Zk prediction}) under $\Delta_1\leftrightarrow \Delta_2$. This orbifold falls into the class studied in Section 5.4.1 of \cite{Dorey:2023jfw}, where their parameters are identified as $p_1^L=1,p_2^L=-1,\alpha_1=0,\alpha_2=1$. } $S^1\times (S^5/\Z_k)$ at large $N$,
\begin{align}
&\log \I_\text{M5}^{(S^5/\Z_k)}(N;\omega_i;\Delta_I) \nn\\
&\quad =  \frac{1}{k}\alpha(\omega_i;\Delta_I)(N^3-1) + \left(\frac{1}{k}\beta(\omega_i;\Delta_I) - \frac{\Delta_2}{24}\left(k-\frac{1}{k}\right)\right)(N-1) + \cO\!\left(\log N\right)\ .
\label{eq: S5/Zk prediction}
\end{align}
It would then be very interesting to match this result to the first few orders in the Cardy expansion of the M5-brane theory.

Furthermore, the 6d index on $S^1\times (S^5/\Z_k)$ then admits a so-called `ultra-spinning' limit, in which it is given by the superconformal index of a supersymmetric $\sigma$-model on the moduli space of Yang-Mills instantons \cite{Dorey:2023jfw}. The dual black hole geometries were constructed and studied in \cite{Dorey:2022cfn,Mouland:2023gcp}. It would be interesting to understand the interplay of this limit with the giant graviton expansion, and in particular the $k>1$ extension of the relation (\ref{eq: main claim}).  

\item A central player in this work has been the superconformal index $\I_\text{M2}(\tN;\tw;\tDelta_A)$ of the ABJM theory. We can however apply similar manipulations to another exactly-computable observable: the supersymmetric partition function $\cZ_{S_b^3}(\tN;\dots)$ on the squashed $3$-sphere $S^3_b$. Can we extract information about the 6d $(2,0)$ theory from the large $\tN$ asymptotics of $\cZ_{S_b^3}(\tN;\dots)$, as we did for the superconformal index? We no longer have a giant graviton expansion to rely on, and so we proceed speculatively.

We specialise to level $k=1$. Then $\cZ_{S_b^3}(\tN;b^2;v_A)$ depends on a squashing parameter $b$ along with four chemical potentials\footnote{One usually uses the symbol $\Delta$ for these chemical potentials, but having used $\Delta_I$ and $\tDelta_A$ relentlessly throughout this paper, we thought we'd give the reader a break and use $v$.} $v_A$ for the $SO(8)$ R-symmetry, which satisfy $\sum_A v_A = -2\pi i$. Much like for the superconformal index, \cite{Hristov:2022lcw,Bobev:2022jte,Bobev:2022eus} proposed an expression for this partition function to all perturbative orders at large $\tN$. It is given by\footnote{Our parameters are related to those of \cite{Hristov:2022lcw} by $b_\text{ours}=b_\text{theirs}$ and $(v_A)_\text{ours} = -\pi i (\Delta_A)_\text{theirs}$. }
\begin{align}
  \cZ_{S_b^3}(\tN;b^2;v_A) \sim e^{D(b^2,v_A)} \, \text{Ai}\left[\frac{\tN-B(-2\pi i b^2,(1+b^2)v_A)}{C(-2\pi i b^2,(1+b^2)v_A)^{1/3}}\right] \left(1 + \cO\!\left(e^{-\sqrt{\tN}}\right)\right)\ ,
  \label{eq: S3 Airy}
\end{align}
where $B(\tw,\tDelta_A),C(\tw,\tDelta_A)$ are given in (\ref{Def:3d-CB}), while $D(b^2,v_A) = D(1/b^2,v_A)$ is unknown. Note that $\cZ_{S_b^3}$, and accordingly the right-hand-side of this expression, are invariant under $b\to 1/b$. One can check that the identification of the parameters $(\tw,\tDelta_A) = (-2\pi i b^2,(1+b^2)v_A)$ in the evaluations of $B(\tw,\tDelta_A),C(\tw,\tDelta_A)$ are precisely such that the constraint (\ref{eq: linear constraint again}) is satisfied.

Indeed, this identification of parameters and the close relationship between (\ref{eq: S3 Airy}) and (\ref{eq: Hristov full})---differing essentially just in the power of the Airy function---can be understood rather cleanly in the 3d Cardy limit $\tw\to 0$. Leveraging the known holomorphic factorisation into two holomorphic blocks of both $\I_\text{M2}$ and $\cZ_{S_b^3}$ \cite{Pasquetti:2011fj, Beem:2012mb}, it was shown in \cite{Choi:2019dfu} that in this limit one has schematically $\log \I_\text{M2}\sim \log \cZ_{S_b^3} + \log \tilde{\cZ}_{S_b^3}$ with parameters identified as above\footnote{The second factor $\tilde{\cZ}_{S_b^3}$ has everything flipped by a sign, i.e. $(\tw,\tDelta_A) = -(-2\pi i b^2,(1+b^2)v_A)$, under which the asymptotic (\ref{eq: S3 Airy}) is invariant. }.

We'd then like to point out two immediate consequences of (\ref{eq: S3 Airy}). The first is that the grand canonical $S^3_b$ partition function $\Xi_{S^3_b}(\mu;b^2;v_A) = \sum_{\tN} e^{\mu \tN} \cZ_{S_b^3}(\tN;b^2;v_A) $ obeys
\begin{align}
  &\log \Xi_{S^3_b}\left( -\Delta_2 N;-\Delta_1/2\pi i;\frac{\omega_i}{1-\Delta_1/2\pi i},\frac{-\Delta_2}{1-\Delta_1/2\pi i}\right) \nn\\
  & \hspace{40mm} \sim 4\alpha(\omega_i;\Delta_I)(N^3-1) + \beta(\omega_i;\Delta_I)(N-1) + \dots \ ,
\end{align}
at large $N$. The right-hand-side differs from the thermal anomaly polynomial of the $U(N)$ 6d $(2,0)$ theory only in the factor of $4$ in the first term.

Secondly, if we instead define $\Xi^\text{double}_{S^3_b}(\mu;b^2;v_A) = \sum_{\tN} e^{\mu \tN} (\cZ_{S_b^3}(\tN;b^2;v_A))^2 $, we have at large $N$
\begin{align}
  &\log \Xi^\text{double}_{S^3_b}\left( -\Delta_2 N;-\Delta_1/2\pi i;\frac{v_i}{1-\Delta_1/2\pi i},\frac{-\Delta_2}{1-\Delta_1/2\pi i}\right) \nn\\
  & \hspace{40mm} \sim \alpha(\omega_i;\Delta_I)(N^3-1) + \beta(\omega_i;\Delta_I)(N-1) + \dots \ ,
\end{align}
where we recognise the right-hand-side as precisely the thermal anomaly polynomial of the $U(N)$ 6d $(2,0)$ theory.

One may interpret these results as a hint towards a giant graviton expansion of the $S^3_b$ partition function. We leave such an interpretation to future work.

\item Ensembles played a significant role in this work; in particular, the key claim (\ref{eq: main claim}) is a relationship between the grand canonical ensemble of the M2-brane theory, and the canonical ensemble of the M5-brane theory.

The role of these two ensembles in holography has recently been discussed in \cite{Gautason:2025plx}. Let us briefly summarise. In the context of AdS$_4$ holography, it was proposed that the canonical partition function (and so in particular the index $\I_\text{M2}$ defined in (\ref{eq: M2 index})) should correspond in the bulk to the exponential of what is referred to as the M5-brane generating functional. This object lacks a formal definition, but it is heuristically thought of as the generating functional of a theory arising from the quantisation of M5-branes in AdS$_4\times S^7$. As already noted in \cite{Gautason:2025plx} for the $\frac{1}{2}$-BPS index, the giant graviton expansion (\ref{eq: GGE}) can be thought of as a manifest realisation of this idea, for it provides an expression for $\I_\text{M2}$ in terms of the BPS states of M5-branes.

Conversely, it is suggested that the grand canonical index $\Xi_\text{M2}$ should be computed in the bulk as the exponential of a corresponding M2-brane generating functional, arising from the quantisation of M2-branes in AdS$_4\times S^7$. In particular, the order $\mu$ term appearing in the asymptotics (\ref{eq:logXi-M2}) should correspond to an eight-derivative correction in the effective supergravity description of this theory. The form (\ref{eq: geometric sum}) for $\Xi_\text{M2}$ does not readily admit such an interpretation; what one really wants is a `dual giant graviton expansion' of $\Xi_\text{M2}$, towards which we make no progress. However, consider the expression (\ref{eq: main claim}) for a particular residue of $\Xi_\text{M2}$. One can then consider expanding $\I_\text{M5}$ in its own giant graviton expansion \cite{Arai:2020uwd}, made up of contributions from wrapped M2-branes in AdS$_7\times S^4$. In this sense, one can in principle leverage (\ref{eq: main claim}) to determine a particular residue of $\Xi_\text{M2}$ in terms of quantised M2-branes, albeit in an AdS$_7\times S^4$ background.

\end{itemize}

\subsection*{Acknowledgments}

The authors would like to thank Yosuke Imamura for many valuable discussions and comments. We would also like to thank Luigi Tizzano and Paul Luis Roehl for helpful discussions, and Jesse van Muiden for comments on the draft.
The work of H.Y.C. was supported in part by Ministry of Science and Technology (MOST) through the grant 
 114-2112-M-002-022-.
The work of S. M.
was supported by JSPS Grant-in-Aid for Scientific Research (C) \#22K03598. 
H. Y. C. also likes to thank Osaka Metropolitan University for the hospitality, where part of this work was carried out. The work of N.D. was partially supported by STFC consolidated grant ST/X000664/1. The work of R.M. was supported by the UK Engineering and Physical Sciences grant EP/Z000106/1. The work of C.\c{S}. was supported by the Czech Academy of Sciences under the grant number LQ100102101, and by European Structural and Investment Funds and the Czech Ministry of Education, Youth and Sports (Project FORTE CZ.02.01.01/00/22 008/0004632).


\appendix 
\addcontentsline{toc}{section}{Appendices}
\addtocontents{toc}{\protect\setcounter{tocdepth}{0}}

\part*{Appendices}

\section{Proof of double $q$-Pochhammer identity}\label{app: indentity proof}

We would like to prove the identity (\ref{eq: identity}). This can be viewed as a generalisation of a well-known identity due to Karpelevich for the $q$-Pochhammer symbol, a proof of which can be found for instance in \cite{olshanetsky1995modifiedqbesselfunctionsqbesselmacdonald}.

\para 
 Let us start by defining the double $q$-Pochhammer symbol
\begin{align}
    (z;p,q)_{r,s} = \prod_{n=0}^{r-1}\prod_{m=0}^{s-1} (1-zp^n q^m)\ .
\end{align}
We can also define
\begin{align}
    (z;p,q)_{\infty,s} &= \prod_{n=0}^{\infty}\prod_{m=0}^{s-1} (1-zp^n q^m)\ ,\qquad |p|<1	\ ,	\nn\\
    (z;p,q)_{r,\infty} &= \prod_{n=0}^{r-1}\prod_{m=0}^{\infty} (1-zp^n q^m)\ ,\qquad |q|<1	\ ,	\nn\\
    (z;p,q)_{\infty,\infty} &= \prod_{n=0}^{\infty}\prod_{m=0}^{\infty} (1-zp^n q^m)\ ,\qquad |p|,|q|<1 \ .
\end{align}
The claimed identity (\ref{eq: identity}) can then be rephrased as the claim that, for $|p|,|q|<1$,
\begin{align}
  \frac{1}{(z;p,q)_{\infty,\infty}} = \sum_{n,m\ge 0}\frac{C_{n,m}}{1-zp^n q^m} \ ,
\label{eq: rephrased theorem}
\end{align}
where
\begin{align}
  C_{n,m} &= \prod_{\substack{u,v \ge 0\\ (u,v)\neq (n,m)}} \frac{1}{1-p^{u-n}q^{v-m}} \nn\\
  &= \frac{1}{(pq;p,q)_{\infty,\infty}((pq)^{-1};p^{-1},q^{-1})_{n,m}(p^{-1};p^{-1},q)_{n,\infty}(q^{-1};p,q^{-1})_{\infty,m}(p;p)_\infty(q;q)_\infty}\ .
\end{align}
Here, $(z;q)_r$ is the usual $q$-Pochhammer symbol,
\begin{align}
  (z;q)_r = \prod_{n=0}^{r-1}(1-zq^n)\ ,
\end{align}
 and
 \begin{align}
  (z;q)_\infty = \prod_{n=0}^{\infty}(1-zq^n),\qquad |q|<1\ .
\end{align}
To prove (\ref{eq: rephrased theorem}), let us first define for each $r,s\in \mathbb{N}$,
\begin{align}
  f_{r,s}(z):=\frac{1}{(z;p,q)_{r,s}} - \sum_{n=0}^{r-1}\sum_{m=0}^{s-1}\frac{C^{r,s}_{n,m}}{1-zp^n q^m}\ ,
\end{align}
where
\begin{align}
  C^{r,s}_{n,m} &= \prod_{\substack{0\le u \le r-1\\0\le v \le s-1\\ (u,v)\neq (n,m)}} \frac{1}{1-p^{u-n}q^{v-m}} \nn\\
  &= \frac{1}{(pq;p,q)_{r-n,s-m}((pq)^{-1};p^{-1},q^{-1})_{n,m}(p^{-1};p^{-1},q)_{n,s-m}(q^{-1};p,q^{-1})_{r-n,m}(p;p)_{r-n}(q;q)_{s-m}}\ .
\end{align}
Then, $f_{r,s}(z)$ is entire on the complex plane, and approaches $f_{r,s}(z)\to 0$ as $|z|\to \infty$, and hence by Liouville's theorem we have $f_{r,s}(z)=0$. So we have
\begin{align}
  \frac{1}{(z;p,q)_{r,s}} = \sum_{n=0}^{r-1}\sum_{m=0}^{s-1}\frac{C^{r,s}_{n,m}}{1-zp^n q^m}\ .
\end{align}
Thus,
\begin{align}
  \frac{1}{(z;p,q)_{\infty,\infty}} = \lim_{r,s\to\infty} \left(\sum_{n=0}^{r-1}\sum_{m=0}^{s-1}\frac{C^{r,s}_{n,m}}{1-zp^n q^m}\right)\ .
\end{align}
We have that
\begin{align}
   \lim_{r,s\to\infty} C^{r,s}_{n,m} = C_{n,m}\ ,
\end{align}
for each fixed $n,m$. We also have
 \begin{align}
 |C^{r,s}_{n,m}| < |C_{n,m}|\ ,
\end{align}
and thus by dominant convergence we have
\begin{align}
  \frac{1}{(z;p,q)_{\infty,\infty}} = \sum_{n=0}^{r-1}\sum_{m=0}^{s-1}\left( \lim_{r,s\to\infty}\frac{C^{r,s}_{n,m}}{1-zp^n q^m}\right) = \sum_{n,m\ge 0}\frac{C_{n,m}}{1-zp^n q^m}\ ,
\end{align}
as required.

\section{Asymptotics from the growth of residues}\label{app: asymptotics}

The aim of this appendix is to determine, starting with the expression (\ref{eq: geometric sum}), how the large $\mu$ asymptotics of $\Xi_\text{M2}(\mu;\tw;\tDelta_A)$ are encoded in the large $\{n_A\}$ asymptotics of $\I_{(n_A)}(\tw;\tDelta_A)$.

\para
We begin with by rewriting (\ref{eq: geometric sum}) slightly into the form
\begin{align}
  \Xi_\text{M2}(\mu;\tilde{\omega};\tilde{\Delta}_A) =  \I_\infty (\tilde{\omega};\tilde{\Delta}_A) \sum_{n_A\in \mathbb{N}_0^4} \frac{ \I_{(n_A)}(\tilde{\omega};\tilde{\Delta}_A)}{   1- e^{\mu - \tilde{\Delta}_A n_A}}\ .
\end{align}
Let us then suppose that at large $\mu$ that the sum is well-approximated by a four-dimensional integral, a step we will later need to justify for self-consistency. Then at large $\mu$ we have
\begin{align}
  \Xi_\text{M2}(\mu;\tilde{\omega};\tilde{\Delta}_A) \sim  \I_\infty (\tilde{\omega};\tilde{\Delta}_A) \int dn_1\dots dn_4\, e^{S(\mu,n_A;\tilde{\omega};\tilde{\Delta}_A)}\ ,
\label{eq: real integral}
\end{align}
where
\begin{align}
  S(\mu,n_A;\tilde{\omega};\tilde{\Delta}_A) = \log \I_{(n_A)}(\tilde{\omega};\tilde{\Delta}_A) - \log \left(1- e^{\mu - \tilde{\Delta}_A n_A}\right)\ .
  \label{eq: saddle action}
\end{align}
The resulting saddle-point equations then read
\begin{align}
  \frac{\partial}{\partial n_B}S(\mu,n_A;\tilde{\omega};\tilde{\Delta}_A) = \frac{\partial}{\partial n_B}\log \I_{(n_A)}(\tilde{\omega};\tilde{\Delta}_A) - \frac{\tDelta_B e^{\mu - \tilde{\Delta}_A n_A} }{1-e^{\mu - \tilde{\Delta}_C n_C}} = 0\ ,
\end{align}
for $B=1,2,3,4$.

\para
Let us in particular assume that $\arg \tDelta_4 \neq \arg \tDelta_{1,2,3}$, and consider the asymptotics of $\Xi_\text{M2}(\mu;\tilde{\omega};\tilde{\Delta}_A)$ as we take $\mu = \tDelta_4 x$ for $x\in \mathbb{R}$, $x\to +\infty$. If we furthermore assume that $\log \I_{(0,0,0,n_4)}(\tilde{\omega};\tilde{\Delta}_A)$ grows like $(n_4)^k$ at large $n_4$ for some $k>1$, we see that at large $\mu$ the four-dimensional real integral (\ref{eq: real integral}) passes infinitesimally close to a saddle-point at
\begin{align}
  n_A = n_A^* = (0,0,0,\mu/\tDelta_4) + \cO\!\left(\frac{1}{\mu^{k-1}}\right)\ .
\label{eq: saddle}
\end{align}
Assuming we can deform the contour slightly to pass through this saddle-point, which indeed lies at large $n_4$, we ultimately find that at large positive real $x$, we have
\begin{align}
  \log \Xi_\text{M2}(\tDelta_4 x;\tilde{\omega};\tilde{\Delta}_A) \sim \log \I_{(0,0,0,x)}(\tilde{\omega};\tilde{\Delta}_A) + \cO(\log x)\ ,
\end{align}
where note that $\log \I_\infty$ gives rise to a constant in $x$ term that is included in the $\cO(\log x)$ correction.
Note that there are several sources of corrections of order $\log x$. Two of these are the evaluation of the second term in (\ref{eq: saddle action}), which gives $(k-1)\log x$; and the 1-loop correction to the saddle-point approximation, which we do not compute. But in addition we also need to consider the size of the error introduced when we took the Euler-Maclaurin approximation (\ref{eq: real integral}). Schematically, this error can be written in the form \cite{EM}\footnote{Note that the expressions appearing in \cite{EM} strictly only apply to integrals over finite polytopes. As is common practise for one-dimensional Euler-Maclaurin approximations, we assume our summand is sufficiently well-behaved that these same expressions can be applied here. Alternatively one can arrive at the expression (\ref{eq: EM remainder}) by repeated use of the one-dimensional Euler-Maclaurin formula, where the operator $D$ can be deduced by keeping track of the relevant remainder terms.}
\begin{align}
    \int dn_1\dots dn_4\, e^{S(\mu,n_A;\tilde{\omega};\tilde{\Delta}_A)} - \sum_{n_1,n_2,n_3,n_4}  e^{S(\mu,n_A;\tilde{\omega};\tilde{\Delta}_A)} = \int dn_1\dots dn_4\, D\!\left(e^{S(\mu,n_A;\tilde{\omega};\tilde{\Delta}_A)}\right)\ ,
    \label{eq: EM remainder}
\end{align}
where $D$ is a partial differential operator in $n_A$ space, i.e. a finite polynomial of $(\partial /\partial n_A)$, $A=1,2,3,4$. Since we assume $S$ grows polynomially with the $n_A$, it follows that the integral appearing here can be approximated at the same saddle (\ref{eq: saddle}), and hence the contribution of this error term to $\log \Xi_\text{M2}(\tDelta_4 x;\tilde{\omega};\tilde{\Delta}_A)$ is no larger than $\log x$.

\para
Turning this logic around, we arrive at the claim that (\ref{eq: relation of asymptotics}) follows from (\ref{eq: asymptotic series}). Note that the key assumption we need is that the asymptotic (\ref{eq: asymptotic series}) holds in a wedge that includes the ray $\mu = \tDelta_4 x$, $x\in \mathbb{R}_{>0}$.




\section{Grand canonical ensemble of superconformal index with fluxes}\label{App:withfluxes}
In this appendix, we consider the M2-brane grand canonical superconformal index with non-vanishing background R-symmetry magnetic fluxes. Although the inclusion of fluxes is not necessary for the present work, it may be interesting to investigate the flux dependence in the future. The extension of the proposal (\ref{eq: Hristov full}) to include fluxes (as well as generic level $k$) is given by \cite{Hristov:2022lcw} 
\begin{equation}
\label{eq:ZSCI-Flux}
\cI_{\text{M2}}(\tN;\tw, \tDelta_A, \bn_A)
\simeq
C_{+}^{-1/3}\mathrm{Ai}\left[C_{+}^{-1/3}(\tN - B_{+})\right]
\times C_{-}^{-1/3}\mathrm{Ai}\left[C_{-}^{-1/3}(\tN - B_{-})\right]\ ,
\end{equation}
where $\bn_A \in \bbZ$ labels quantized background R-symmetry fluxes. In this appendix we use $\simeq$ as in \cite{Hristov:2022lcw} to denote equality up to an overall $\tN$-independent constant, and non-perturbative corrections. This expressions is written in terms of
\begin{equation}\label{eq:CB-SCI}
C_{\pm} = \frac{2 \tw^2}{\pi^2 k \prod_A (\tDelta_A \pm \tw \bn_A)}\ , 
\quad
B_{\pm} = \frac{k}{24} 
+ \frac{ (\tw + 1)^2 k_{\mathbb{T}}(\tDelta \pm \tw \bn) 
+ (\tw - 1)^2 k_{\mathbb{W}}(\tDelta \pm \tw \bn) }
{48 k \prod_A (\tDelta_A \pm \tw \bn_A)}\ ,
\end{equation}
and the functions $k_{\mathbb{T}}(\tDelta)$ and $k_{\mathbb{W}}(\tDelta)$,
\begin{align}
k_{\mathbb{T}}(\tDelta) &:= \sum_{A=1}^4 (\tDelta_A)^2 - \frac{(\tDelta_1 + \tDelta_2 - \tDelta_3 - \tDelta_4)(\tDelta_1 - \tDelta_2 + \tDelta_3 - \tDelta_4)(\tDelta_1 - \tDelta_2 - \tDelta_3 + \tDelta_4)}{\sum_{A=1}^4 \tDelta_A}\ , \nn\\
k_{\mathbb{W}}(\tDelta) &:= -2 \sum_{A<B} \tDelta_A \tDelta_B\ .\label{eq:kTkW2}
\end{align}
Notice in contrast with the main text, the normalisation of the 3d fugacity parameters have been restored to those in \cite{Hristov:2022lcw} such that $\tDelta_A \to i\frac{\tDelta_{A}}{\pi}$ and $\tw \to i\frac{\tw}{2\pi}$ and $\sum_{A =1}^4 \tDelta_A = 2(1+\tw)$.
The result for the grand canonical partition function is given as below.
\begin{equation}\label{eq:Xi-SCI-exp}
\Xi_{\rm{M2}}(\mu;\Delta_i, \bn_i, \omega) \simeq 
\sum_{\ell \in \bbZ} \frac{(C_+ C_-)^{\frac{1}{3}}}{|\Delta C|^{\frac{1}{3}}}
\mathrm{Ai}\left(\frac{\Delta B \Delta C + C_+ C_- (\mu+2i\pi\ell)^2 }{|\Delta C|^{\frac{4}{3}}}\right)\exp(i\Phi(\mu+2i\pi \ell))\ ,  
\end{equation}
where we have introduced
\begin{equation}
\Delta B = B_+- B_-\ , \quad \Delta C = C_+ - C_-\ ,   
\end{equation}
and
\begin{equation}\label{eq:Phi}
i\Phi(\mu) = \frac{1}{3}\frac{C_+ C_- (C_+ + C_-)}{|\Delta C|^2}\mu^3+ \frac{(C_-B_+ - C_+B_-)}{|\Delta C|}\mu\ .
\end{equation}

\section{Thermal anomaly polynomial and supersymmetric Schur polynomials}\label{App:superschur}
In this appendix, for completeness, we include the general form of the thermal polynomial $P_8^T$ for a 6d $\cN=(2,0)$ $G=ADE$ theory and its equivariant integral following \cite{Nahmgoong:2019hko}.
This was defined originally through the thermal partition function on $S^1 \times_\beta S^5$ with $\beta =\frac{2\pi r_1}{r_5}$, and can be computed using the equivariant integral of the thermal anomaly polynomial $P_8^T$. 
The final expression is obtained from the ordinary anomaly polynomial $P_8$ given in \cite{Ohmori:2014kda} by including the fictitious background gauge field $\bA^{\rT}$ associated with the gravi-photon along the thermal $S^1$ using the following replacement rule,
\begin{eqnarray}\label{eq:P8T-P8 relation}
P_{8}^{T} &=& P_8 \left(p_k(\rT) \to p_{k}(\rT)-\frac{\bF^2_{\rm T}}{4\pi^2}p_{k-1}(\rT)\right) \nonumber\\
&=& 
\frac{h_G^\vee d_G}{24}\, p_2(\rN)
+
\frac{r_G}{48}
\Bigg[
p_2(\rN)
-
\Big(
p_2(\rT)
-
\frac{\mathbf{F}_{\rm{T}}^2}{4\pi^2} p_1(\rT)
\Big)
+
\frac{1}{4}
\Big(
p_1(\rN) - p_1(\rT) + \frac{\mathbf{F}_{\rm{T}}^2}{4\pi^2}
\Big)^2
\Bigg]\ ,\nn\\
\end{eqnarray}
where $p_k(\rN)$ and $p_{k}(\rT)$ are the $k$-th Potryagin class of the normal bundle and tangent bundle, and $\bF_{\rm T}=d\bA_{\rm T}$ is the field strength for $\bA_{\rm T}$. The equivariant integral for $P_8^{T}$ which differs from the one computed in \cite{Bobev:2015kza} using $P_8$ is
\begin{equation}\label{eq:General SCE }
 -\int_{(\omega_i, \Delta_I)} P_8^{T}=
{\alpha}(h^\vee_G d_G + r_G )+{\beta} r_G\ ,
\end{equation}
where 
\begin{align}
    {\alpha}&=-\frac{1}{384\w^{(3)}} \left( \Delta_+^2 - \Delta_-^2 \right)^2\ , \nn\\
    {\beta}&=-\frac{1}{192\w^{(3)}}\Big[
{ \left( \Delta_+^2 + 4\pi^2 \right) \left( \Delta_-^2 + 4\pi^2 \right)}
- (\Delta_+^2 + \Delta_-^2 - 8\pi^2 ) \left((\w^{(1)})^2-2 \w^{(2)}\right)\nonumber\\
&\hspace{25mm}+\left((\w^{(1)})^4-4 (\w^{(1)})^2 \w^{(2)}+8 \w^{(1)} \w^{(3)}\right) 
\Big]\ ,
\label{tildeB}
\end{align}
and $h^{\vee}_G, d_G$ and $r_G$ are the dual Coxeter number, dimension and rank of $G$, respectively. 
We have also defined $\Delta_{\pm} = \Delta_1\pm \Delta_{2}$ and used the notation $\w^{(n)}$ as defined in \eqref{eq:omega-n}. 
If we focus on $G=A_{N-1}\cong SU(N)$, then $d_{A_{N-1}} =N^2-1$, $ h^{\vee}_{A_{N-1}}=N$, $r_{A_{N-1}} = N-1$, while for $G=D_N \cong SO(2N)$, we have $d_{D_{N}} = N(2N-1)$, $ h^{\vee}_{D_{N}}=2N-2$, $r_{D_{N}} = N$. As we see in the following, $\alpha$ and $\beta$ coincide with $\alpha(\w_i, \Delta_I)$ and $\beta(\w_i, \Delta_I)$  defined in \eqref{eq: alpha and beta} under the 6d supersymmetric constraint \eqref{eq: M5 linear constraint}. 
\paragraph{}
Here we also want to point out a hidden structure reminiscent of supersymmetric Schur polynomials for the numerator of the coefficient $\beta$ \eqref{tildeB} of the supersymmetric Casimir energy \eqref{eq:General SCE }.
For this purpose, let us first note that the numerator is homogeneous in various chemical potentials if we regard the constant $\pi$ also as a variable.
Then, we can rewrite the numerator by eliminating the constant $\pi$ using the constraint \eqref{eq: M5 linear constraint} and reproduce \eqref{eq: alpha and beta},
\begin{align}
-24\w^{(3)}\beta&=\w^{(1)}\w^{(3)}-\w^{(1)}\w^{(2)}(\Delta_1+\Delta_2)+ (\w^{(1)})^2\Delta_1\Delta_2+\w^{(2)}(\Delta_1^2+\Delta_2^2+\Delta_1\Delta_2)\nn\\
&\qquad -\w^{(1)}\Delta_1\Delta_2(\Delta_1+\Delta_2)\ .
\end{align}
Naively, even if we formally change variables by $\omega_4=-\Delta_2$, the numerator does not have to be a symmetric function of four chemical potentials for angular momentum including $\omega_4$.
However it turns out to be the case.
If we introduce elementary symmetric polynomials including $\omega_4$, $ e_1'=\sum_{i=1}^4\omega_i$, $e_2'=\sum_{i<j}^4\omega_i\omega_j$, $ e_3'=\sum_{i<j<k}^4 \omega_i\omega_j\omega_k$ and $ e_4'=\omega_1\omega_2\omega_3\omega_4$, the numerator turns out to be given simply by
\begin{align}
-24\w^{(3)}\beta=e_1' e_3'- e_4'- e_1' e_2'\Delta_1+ e_2'\Delta_1^2\ .
\end{align}
Surprisingly, we can continue this rewriting with one more step.
Namely, if we formally change variables by $\omega_5=-\Delta_1$, and introduce elementary symmetric polynomials including $\omega_4$ and $\omega_5$, $e_1''=\sum_{i=1}^5\omega_i$, $e_3''=\sum_{i<j<k}^5 \omega_i \omega_j \omega_k$ and $e_4''=\sum_{i<j<k<l}^5 \omega_i\omega_j\omega_k\omega_l$, the numerator is given simply by
\begin{align}
-24\w^{(3)}\beta=e_1'' e_3''-e_4''\ ,
\end{align}
or in terms of corresponding Schur polynomials, $s''_{2,1,1}$.

\para 
All of this rewriting is reminiscent of the so-called cancellation property of supersymmetric Schur polynomials, whereby the substitution of $x_i=z$ and $y_j=-z$ for the supersymmetric Schur polynomial $s_\lambda(x|y)$ yields a polynomial independent of $z$.

\bibliography{M2M5-GG}
\bibliographystyle{JHEP}

\end{document}